\documentclass[%
reprint,
superscriptaddress,
amsmath,amssymb,
aps,
]{revtex4-2}

\usepackage{graphicx}
\usepackage{float}
\usepackage{dcolumn}
\usepackage{bm}
\usepackage[T1]{fontenc}
\usepackage[utf8]{inputenc}
\usepackage[hidelinks]{hyperref}

\DeclareUnicodeCharacter{202F}{\,}
\newcommand\numberthis{\addtocounter{equation}{1}\tag{\theequation}}
\usepackage{xcolor}

\begin{document}
	
	\preprint{APS/123-QED}
	
	\title{Large impact of phonon lineshapes on the superconductivity of solid hydrogen}
	
	\author{{\DJ}or{\dj}e Dangi{\'c}}
	\email{dorde.dangic@ehu.es}
	\affiliation{Fisika Aplikatua Saila, Gipuzkoako Ingeniaritza Eskola, University of the Basque Country (UPV/EHU), 
		Europa Plaza 1, 20018 Donostia/San Sebasti{\'a}n, Spain}
	\affiliation{Centro de F{\'i}sica de Materiales (CSIC-UPV/EHU), 
		Manuel de Lardizabal Pasealekua 5, 20018 Donostia/San Sebasti{\'a}n, Spain}	
	\author{Lorenzo Monacelli}%
	\affiliation{Theory and Simulation of Materials (THEOS), École Polytechnique Fédérale de Lausanne, CH-1015 Lausanne, Switzerland}
	\author{Raffaello Bianco}
	\affiliation{Ru\dj er Bo\v{s}kovi\'c Institute, 10000 Zagreb, Croatia}
	
	\affiliation{Dipartimento di Scienze Fisiche, Informatiche e Matematiche, Universit\`a di Modena e Reggio Emilia, Via Campi 213/a I-41125 Modena, Italy}
	
	\affiliation{Centro S3, Istituto Nanoscienze-CNR, Via Campi 213/a, I-41125 Modena, Italy}
	\author{Francesco Mauri}
	\affiliation{University of Rome, "Sapienza", Dipartimento di Fisica, 
		Piazzale Aldo Moro 5, 00185, Rome, Italy}
	\author{Ion Errea}%
	\affiliation{Fisika Aplikatua Saila, Gipuzkoako Ingeniaritza Eskola, University of the Basque Country (UPV/EHU), 
		Europa Plaza 1, 20018 Donostia/San Sebasti{\'a}n, Spain}
	\affiliation{Centro de F{\'i}sica de Materiales (CSIC-UPV/EHU),
		Manuel de Lardizabal Pasealekua 5, 20018 Donostia/San Sebasti{\'a}n, Spain}
	\affiliation{Donostia International Physics Center (DIPC),
		Manuel de Lardizabal Pasealekua 4, 20018 Donostia/San Sebasti{\'a}n, Spain}

	\date{\today}
	
	\begin{abstract} 
		Phonon anharmonicity plays a crucial role in determining the stability and vibrational properties of high-pressure hydrides. Furthermore, strong anharmonicity can render phonon quasiparticle picture obsolete questioning standard approaches for modeling superconductivity in these material systems. In this work, we show the effects of non-Lorentzian phonon lineshapes on the superconductivity of high-pressure solid hydrogen. We calculate the superconducting critical temperature T$_\mathrm{C}$ \emph{ab initio} considering the full phonon spectral function and show that it overall enhances the T$_\mathrm{C}$ estimate. The anharmonicity-induced phonon softening exhibited in spectral functions increases the estimate of the critical temperature, while the broadening of phonon lines due to phonon-phonon interaction decreases it. Our calculations also reveal that superconductivity emerges in hydrogen in the $Cmca-12$ molecular phase VI at pressures between 450 and 500 GPa and explain the disagreement between the previous theoretical results and experiments. 
	\end{abstract}
	
	\maketitle
	
	
	\noindent
	\textbf{Introduction.}\\
	Solid atomic hydrogen was postulated to be a high-temperature superconductor at high pressures by Ashcroft in 1968~\cite{Ashcroft1968}. Later this idea has been revised and hydrogen-rich compounds have been hypothesized to be high-temperature superconductors at pressures that are only a fraction of the one needed to get atomic hydrogen~\cite{Ashcroft2004,H3Spred}. The first experimental verification of that idea came in 2015 when H$_3$S was shown to have a transition temperature of 203 K at 155 GPa~\cite{H3S}. This has been followed up by numerous experiments on different hydrogen compounds, many of them exhibiting high-temperature superconductivity~\cite{LaH101,LaH102,YH,CaH,YH2,ThH,PhysRevLett.126.117003}, verifying without a reasonable doubt the existence of superconductivity in hydrides at high pressures~\cite{Eremetsresponse}.  
	
	The discovery of high-temperature superconductivity renewed the interest in synthesizing atomic metallic hydrogen, which is expected to superconduct above room temperature~\cite{supH2,Yan20111264,IonH,Maksimov2001569}. Recently, a work reported atomic metallic hydrogen at 495 GPa on the basis of enhanced optical reflectivity~\cite{H495}. While this finding was questioned~\cite{H495v2} due to a probable overestimation of the measured pressure, there is an abundant amount of proof of finite electrical conductivity of solid hydrogen in the molecular phase~\cite{H360, H420}. None of these works, however, observed the transition to the superconducting phase up to 440 GPa~\cite{PressureScale}. Many first-principles calculations predict the onset of superconductivity in solid hydrogen at significantly lower pressures~\cite{DoganC2c, DoganH, IonCmca}. The disagreement with experiments in this case is surprising in light of the success of the first-principles approach to superconductivity in other high-pressure hydrides~\cite{H3Spred, IonLaH, IonPdH}.
	
	A better understanding of the high-pressure solid hydrogen phase diagram was provided by recent first-principles calculations considering both electronic correlations beyond density functional theory (DFT) and nuclear quantum effects~\cite{Monacellimetal, MonacelliPD, HRQMC}. Monacelli et al. show that at pressures lower than 422 GPa hydrogen crystallizes in the $C2/c-24$ phase, with 24 atoms in the primitive unit cell (phase III of solid hydrogen). In a pressure range between 422 and 577 GPa hydrogen transforms to the $Cmca-12$ phase, with 12 atoms per unit cell (phase VI). The value of 422 GPa agrees very well with the experimental transition pressures detected by infrared at 420 GPa~\cite{H420} and by Raman at 440 GPa~\cite{H360}. Finally, at pressures higher than 577 GPa, hydrogen transforms into atomic hydrogen with a tetragonal $I4_1/amd-2$ structure, containing two atoms per primitive unit cell. 
	
	One of the key reasons why studies in Refs.~\cite{Monacellimetal, MonacelliPD} were able to successfully model the phase diagram of solid hydrogen was the inclusion of quantum anharmonic effects. The phonon renormalization due to anharmonicity can significantly alter superconductivity, as shown in Refs.~\cite{IonHS, IonHS2, IonPdH, DoganH, DoganC2c}. However, these studies have not explored the anharmonicity-induced dynamical renormalization of phonons and its impact on superconductivity. Some studies have highlighted the importance of these effects on superconductivity utilizing simple single phonon mode toy models~\cite{Zaccone1, Zaccone2}. On the other hand, dynamical renormalization of phonons due to electron-phonon coupling has been shown to have little impact on the critical temperature~\cite{Girotto} of conventional superconductors. However, the dynamical effects due to phonon-phonon interaction should be much stronger in high-pressure hydrides, and thus a full first principle study of these effects is necessary.   
	
	Here we present a first-principles study of the superconducting properties of solid hydrogen in its high-pressure phases from 300 to 600 GPa by accounting for quantum anharmonic effects both on the phonons and the structure with the stochastic self-consistent harmonic approximation (SSCHA) at zero Kelvin including the dynamical renormalization of phonon quasiparticles~\cite{supp_mat}. We find that the SSCHA appreciably changes the structure of solid hydrogen in all phases, which leads to an increased density of states (DOS) at the Fermi level and an overall phonon softening. These two effects combine to increase the electron-phonon coupling constants and superconducting transition temperatures in the SSCHA structures, at odds with previous calculations that neglect the impact of ionic quantum effects on the structure~\cite{DoganC2c,DoganH}. We also show that the phonon spectral functions of all these phases have a complex and broad shape, clearly deviating from a simple Lorentzian, questioning the standard approximation made in the electron-phonon calculations in which the spectral function is represented with a Dirac delta function. By considering the full phonon spectral function, we show that the critical temperature (T$_{\text{C}}$) of both molecular and atomic phases is considerably enhanced.  Our calculations predict the onset of superconductivity in solid hydrogen in the semimetallic molecular phase VI at pressures between 450 and 500 GPa, which is consistent with recent experiments~\cite{H360}. 
	
	\begin{figure}[t]
		\begin{center}
			\includegraphics[width=0.95\linewidth]{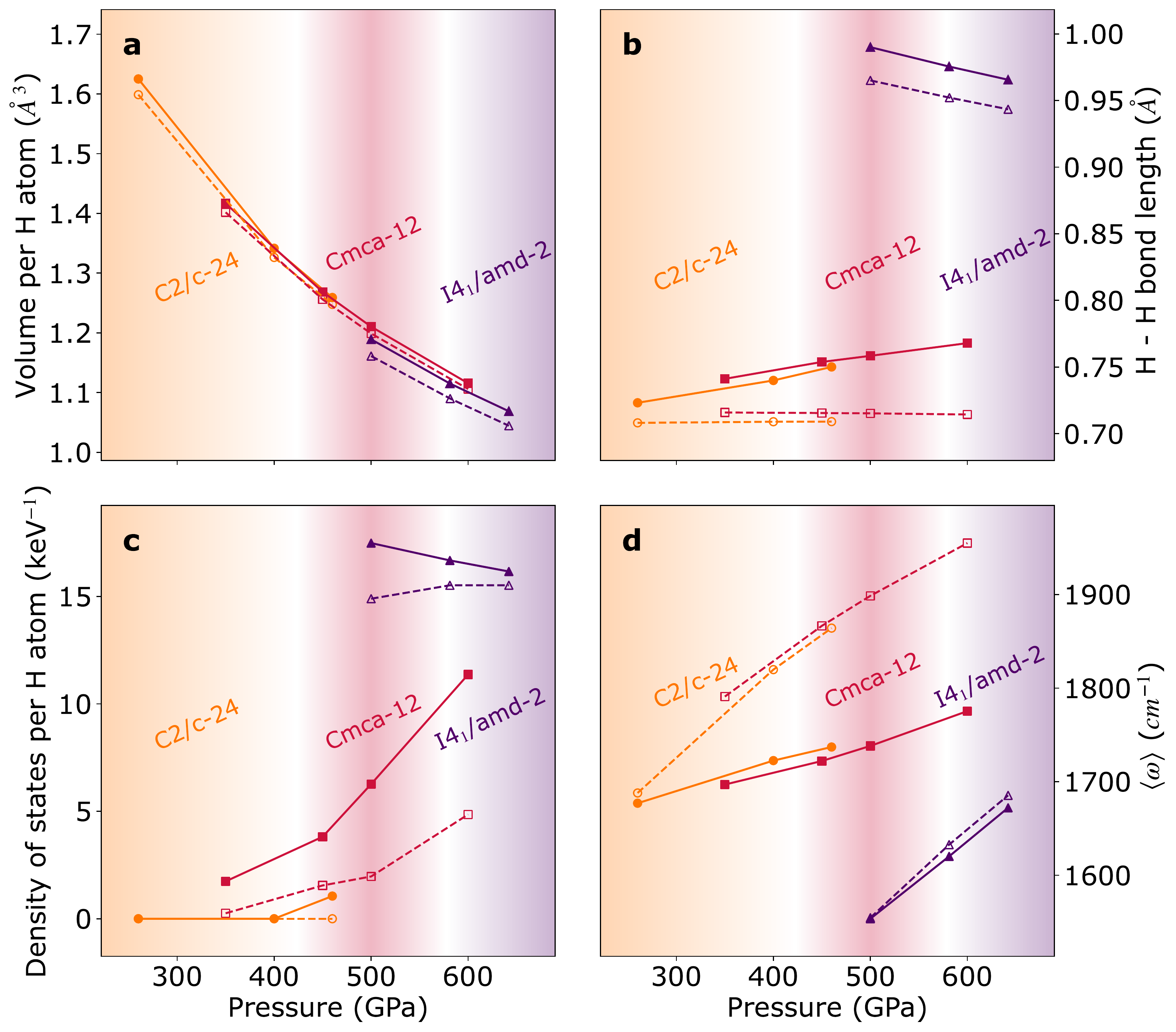}
			\caption{\textbf{a} Volume of the primitive unit cell per hydrogen atom, \textbf{b} length of the hydrogen-hydrogen bond, \textbf{c} the electronic DOS at the Fermi level per hydrogen atom, and \textbf{d} the average phonon frequency in high-pressure solid hydrogen. Solid lines represent data obtained for structures relaxed within SSCHA considering quantum anharmonic effects and dashed lines are for the structures that are minima of the Born-Oppenheimer energy surface. The color background shows a phase diagram of the solid hydrogen from Ref.~\cite{MonacelliPD} and the color of the lines indicates for which phase calculations were performed.}
			\label{fig1}
		\end{center}
	\end{figure}
	
	\begin{figure}[!t]
		\begin{center}
			\includegraphics[width=0.95\linewidth]{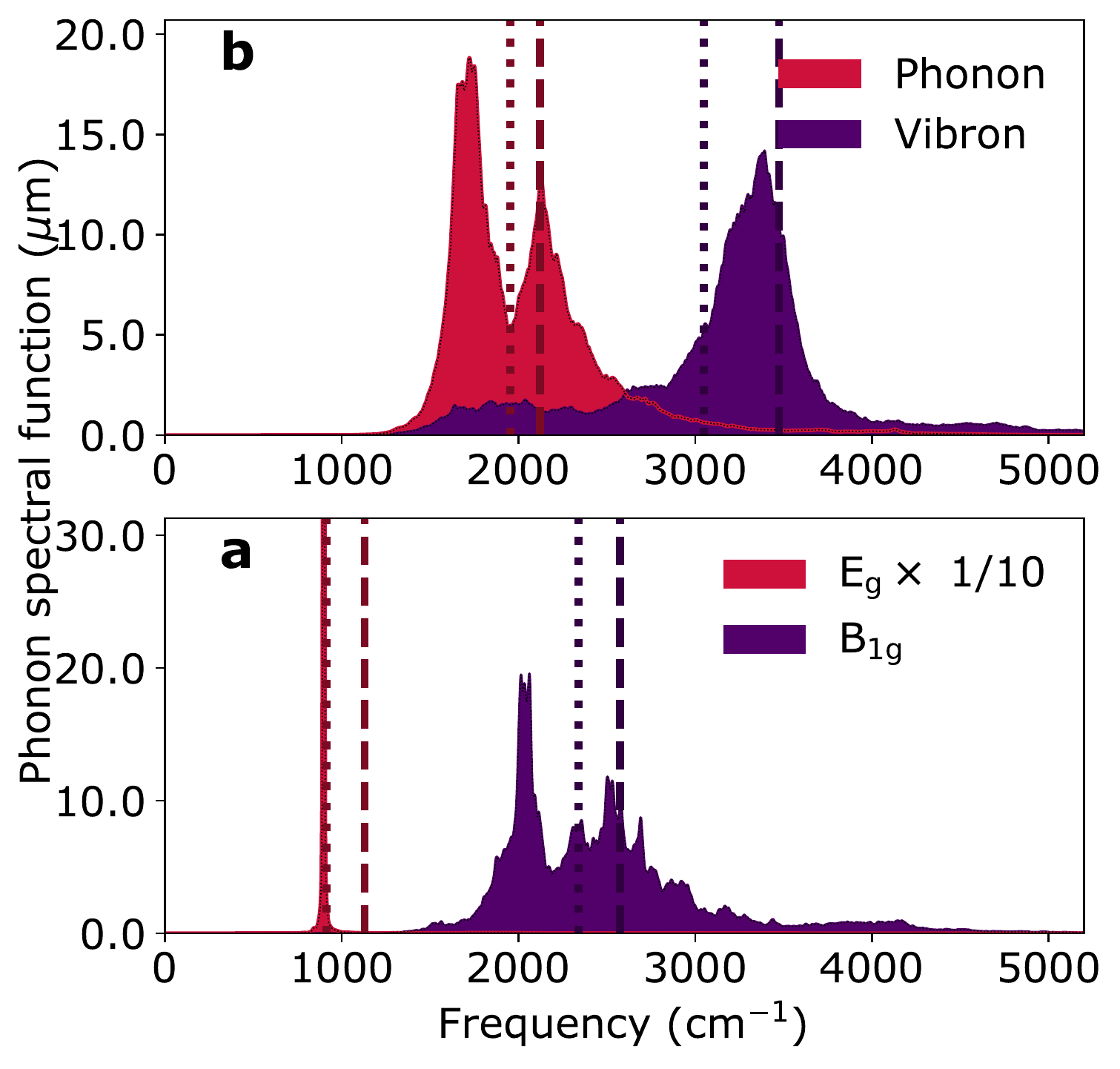}
			\caption{Phonon spectral functions in the no mode mixing approximation in mode basis, $\sigma_{\mu}(\mathbf{q},\omega)$, of two representative optical phonon modes at $\Gamma$ of solid hydrogen in \textbf{a} molecular $Cmca-12$ phase VI at 500 GPa, and \textbf{b} atomic tetragonal $I4_1/amd-2$ phase at 500 GPa. In figure \textbf{b} we scaled the values of the E$_\mathrm{g}$ mode in order to make the figures clearer. Thick dashed vertical lines represent the corresponding frequencies obtained from the auxiliary SSCHA force constants, while dotted lines represent the corresponding free energy Hessian frequency.}
			\label{fig2}
		\end{center}
	\end{figure}

	\noindent
	\textbf{Results and discussion.}\\
	Quantum anharmonic effects have a large impact on the structures in the phase diagram as shown in Fig.~\ref{fig1} (solid lines), compared to the structures that are minima of the Born-Oppenheimer energy surface (BOES) (dashed lines). There is a discontinuity in volume at the phase transition between molecular and atomic phases, not evident for the transition between molecular phases III and VI. This discontinuity is partly suppressed in the quantum anharmonic SSCHA structures. The SSCHA expands the structure slightly for all phases, most prominently for the atomic phase, increasing bond lengths and the $c/a$ ratio at all pressures, as it has been already calculated in other high-pressure hydrides~\cite{HouSc6, IonCmca}. Importantly, SSCHA changes the qualitative behavior of bond lengths in molecular phases: while in SSCHA the bond length increases with pressure, in the classical harmonic approximation, in which it is determined by the minimum of the BOES, it stays relatively constant~\cite{Monacellimetal}. 
	
	These changes have a significant effect on the electronic and vibrational properties of solid hydrogen (see Figs.~\ref{fig1} and ~\ref{fig2}). The most prominent impact is the increase of the DOS at the Fermi level in the quantum anharmonic SSCHA structures. 
	In the molecular phase VI, decreasing volume leads to an increase in the DOS, but with a  considerably higher slope for the SSCHA structures than for the harmonic ones. This behavior shows that quantum anharmonic effects tend to increase the DOS at the Fermi level, as already described in several hydrides~\cite{IonLaH,IonCmca}. 
	Molecular phase III is only weakly semimetallic up to 450 GPa and will not be discussed further on, as, thus, it cannot superconduct as suggested by the latest transport experimental results~\cite{H360}. Closing of the fundamental band gap in our DFT calculations occurs above 400 GPa, which is slightly overestimated compared to calculations that include both better approximation for the exchange-correlation functional and the effect of the electron-phonon coupling~\cite{Monacellimetal, HRQMC, Rillo}. The effects of the electron-phonon coupling (which is the main driver of the band gap closure) will be somewhat included in our superconductivity calculations through the self-consistent solution of Eliashberg equations.
	
	In addition to the structure modified by quantum nuclear effects, the SSCHA method allows us to obtain auxiliary second-order force constants renormalized by anharmonicity. Quantum anharmonicity softens phonon frequencies as a consequence of the stretching of the H bonds (see Fig.~\ref{fig1}). This is at odds with recent calculations~\cite{DoganC2c,DoganH}, in which the frequencies of the phonon modes excluding the vibrons increase due to anharmonicity. The difference is that, in the latter case, the effect of the quantum zero-point fluctuations on the structure was neglected, which our calculations show to be important. Additionally, in the self-consistent harmonic approximation of Ref.~\cite{DoganH} a truncated potential is used (to the fourth order), which gives slightly different results compared to the SSCHA method where all anharmonic orders are included in the calculation of the auxiliary force constants. Both the increase of the DOS at the Fermi level and the phonon softening are beneficial for superconductivity since the electron-phonon coupling constant scales inversely with phonon frequencies and linearly with the DOS at the Fermi level. 
	
	\begin{figure}[!t]
		\begin{center}
			\includegraphics[width=0.95\linewidth]{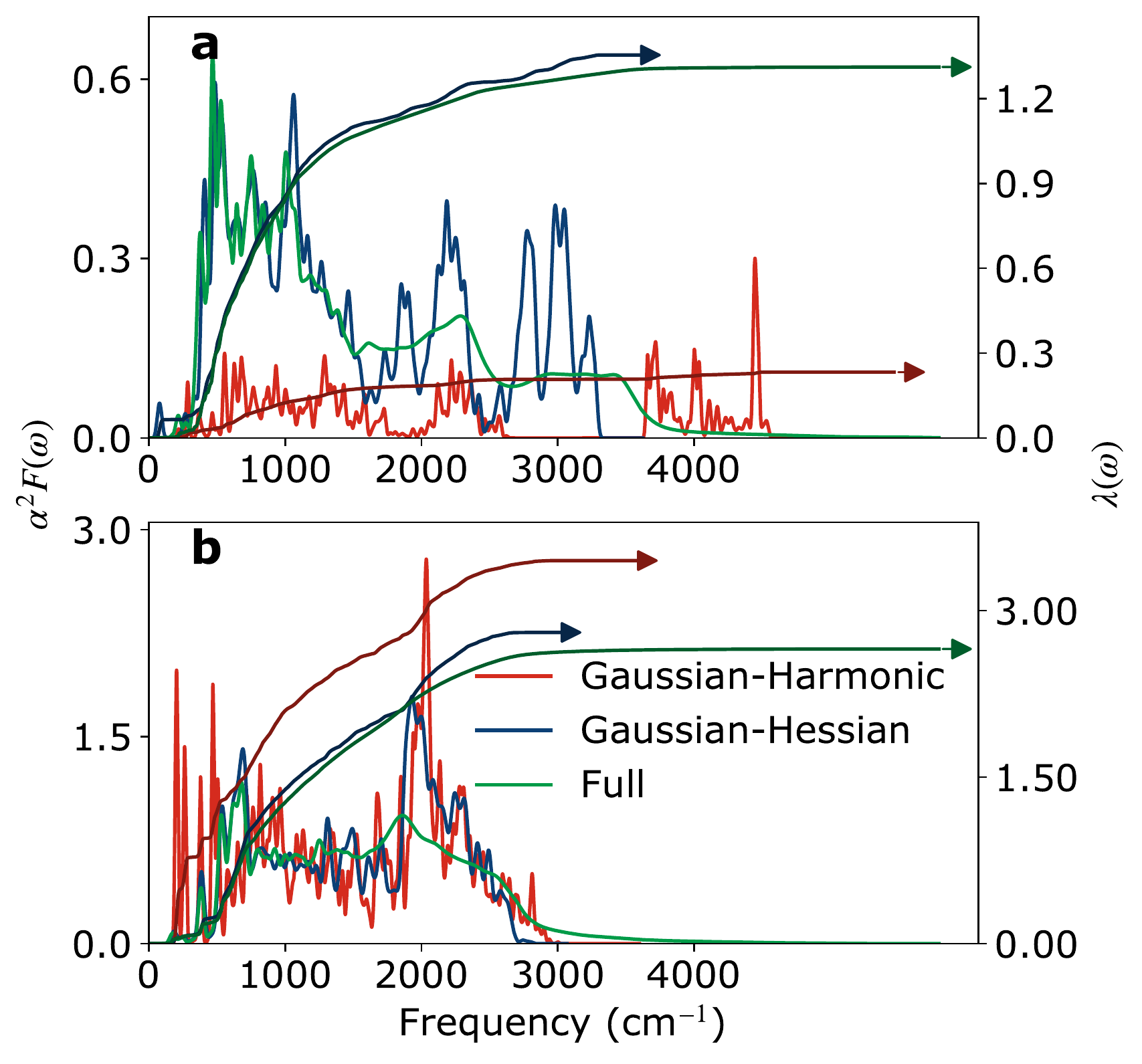}
			\caption{Eliashberg spectral function $\alpha^2F(\omega)$ and integrated electron-phonon coupling constant $\lambda(\omega)$ of solid hydrogen in \textbf{a} molecular $Cmca-12$ phase VI at 500 GPa, and \textbf{b} atomic tetragonal $I4_1/amd-2$ phase at 500 GPa. The "Gaussian-Harmonic" label refers to results calculated with harmonic phonons of the DFT structures in the Gaussian approximation for the phonon spectral function. "Gaussian-Hessian" refers to results calculated with phonons from free energy Hessian calculated for SSCHA structures in the Gaussian approximation for phonon spectral function. Finally, "Full" results were obtained for SSCHA structures with the $\alpha^2F$ calculated with the full phonon spectral function matrix.}
			\label{fig3}
		\end{center}
	\end{figure}
	
	Beyond the renormalization of structural parameters and phonon frequencies, anharmonicity has a huge impact on the phonon spectral function (see Supplementary Material for more details~\cite{supp_mat}). 
	The spectral function of all phases shows further softening with respect to the auxiliary SSCHA phonon frequencies, especially for high-frequency optical modes. This softening can be also captured with the calculation of the free energy Hessian. Specifically, in the static limit, the peaks of the phonon spectral function coincide with the frequencies obtained diagonalizing the free energy Hessian. However, Fig.~\ref{fig2} clearly demonstrates the range of applicability of the free energy Hessian for describing vibrational properties. It is a good approximation in the vicinity of the vanishing imaginary self-energy, that is when auxiliary SSCHA frequency is close to 0, or when there is no large broadening of the phonon spectral line.
	
	In addition to the aforementioned softening, we predict a huge broadening of the phonon spectral functions of the order of thousands of $cm^{-1}$ even at vanishing temperatures. In this case, phonon spectral functions clearly deviate from the standard Lorentzian line shape. We illustrate this in Fig.~\ref{fig2}, where phonon spectral functions for selected modes at $\Gamma$ point are presented for structures at 500 GPa in molecular phase VI and atomic phase. We report two representative modes for molecular phase VI: a global lattice vibration (phonon mode) and a stretching of H$_2$ molecule (vibron mode). In the atomic phase, we only have two optical modes that are non-degenerate and we show both of them. The shift of the phonon frequency is very large in all cases. Additionally, all modes, except the E$_{\mathrm{g}}$ one in the atomic phase, have a huge broadening of the phonon spectral function of thousands of cm$^{-1}$ and a clear non-Lorentzian line shape. Such anomalous behavior questions the standard practice of approximating the spectral function with slightly smeared Delta functions in first-principles calculations of the superconducting critical temperatures. In fact, it has already been shown that non-Lorentzian lineshapes can have a non-negligible effect on other properties of materials, i.e. the lattice thermal conductivity in highly anharmonic semiconducting chalcogenides~\cite{LTC}.
	
	The isotropic Eliashberg function of the electron-phonon interaction can be calculated keeping the full anharmonic spectral function as~\cite{AllenMitrovic}
	\begin{align}
	\alpha^2F(\omega) = \frac{1}{N_{\mathbf{q}}}\sum _{ab\mathbf{q}} \frac{\Delta ^{ab}(\mathbf{q})\sigma_{ab}(\mathbf{q},\omega)}{\omega\sqrt{m_am_b}},
	\label{eq:a2f}
	\end{align}
	where $\sigma_{ab}(\mathbf{q},\omega)$ is the phonon spectral function in the Cartesian basis with wave number $\mathbf{q}$ (see Supplementary Material for more details~\cite{supp_mat}). In Eq. \eqref{eq:a2f} $a$ and $b$ label both atoms and a Cartesian direction,
	$\Delta ^{ab}(\mathbf{q})$ represents the average of the deformation potential over the Fermi surface, $m_a$ is the mass of atom $a$, and $N_{\mathbf{q}}$ is the number of $\mathbf{q}$ points in the sum. In the harmonic case, $\alpha ^2F(\omega)$ is calculated for the structure that minimizes the BOES, while in the SSCHA it is calculated for the structure that minimizes the free energy. Eq. \eqref{eq:a2f} offers a straightforward approach to study the impact of anomalous phonon lineshapes into superconducting properties. However, $\Delta ^{ab}(\mathbf{q})$ includes only the linear term in the electron-phonon interaction without considering higher-order terms that may become important due to quantum nature of hydrogen ions and which are included in other approaches~\cite{bianco,Chen2022Stochastic}.
	
	All calculations thus far that have accounted for anharmonicity in the calculation of $\alpha^2F(\omega)$ have been performed assuming that $\sigma_{ab}(\mathbf{q},\omega)$ can be expressed as \cite{IonPdH,IonH,IonHS,IonHS2,IonLaH,IonCmca}  $\sigma_{ab}(\mathbf{q},\omega)=\sum_{\mu}e^a_{\mu}(\mathbf{q})e^{b*}_{\mu}(\mathbf{q}) \sigma^{\mathrm{h}}_{\mu}(\mathbf{q},\omega)$, where the harmonic spectral function $\sigma^{\mathrm{h}}_{\mu}(\mathbf{q},\omega)$ of mode $\mu$ and wave number $\mathbf{q}$ is a Delta function centered at the harmonic or SSCHA auxiliary phonon frequency, and $\mathbf{e}_{\mu}(\mathbf{q})$ are either harmonic or SSCHA phonon eigenvectors. As in practical implementations, the Delta functions are numerically approximated with a Gaussian function of fixed spread, we label this approach as \emph{Gaussian}. However, as we have shown in Fig.~\ref{fig2}, anharmonicity can drastically affect the phonon lineshapes. In order to obtain $\sigma_{ab}(\mathbf{q},\omega)$, here we utilize the full phonon spectral function. In this case, we do not assume that the phonon self-energy is diagonal in the phonon branch index, as it is done usually, and instead calculate the spectral function as $\sigma_{ab}(\mathbf{q},\omega)=\sum_{\mu\nu}e^a_{\mu}(\mathbf{q})e^{b*}_{\nu}(\mathbf{q}) \sigma_{\mu\nu}(\mathbf{q},\omega)$ fully accounting for off-diagonal terms in phonon self-energy (see Supplementary Material~\cite{supp_mat}). Here the polarization vectors are obtained from the SSCHA auxiliary dynamical matrices. Including full phonon spectral functions drastically changes the calculated $\alpha ^2F(\omega)$, as shown in Fig.~\ref{fig3}. The previously mentioned softening of the phonon modes is also evident in the Eliashberg spectral functions. Additionally, the broadening of the phonon lineshapes leads to the complete closing of the gap between hydrogen vibron and phonon branches in the molecular phase VI. The softening of the phonon modes in the SSCHA coupled with a higher DOS at the Fermi level in the SSCHA structures leads to higher values of the electron-phonon coupling constant $\lambda$ in most cases compared to the harmonic result, more remarkably in the molecular phase VI (see Supp. Material). A notable exception is atomic hydrogen at 500 GPa (depicted in Fig.~\ref{fig3} \textbf{b}), where the proximity to a  phonon instability, which is suppressed by anharmonicity, drastically increases $\lambda$ in the harmonic approximation. Finally, it is worth noting that the no-mode-mixing approximation (treating phonon self-energy as diagonal in phonon branches), which is more commonly used for the calculation of phonon spectral functions, yields similar results to those obtained with the full off-diagonal spectral function (see Supplementary Material~\cite{supp_mat}).

\begin{figure}[t]
		\begin{center}
			\includegraphics[width=0.95\linewidth]{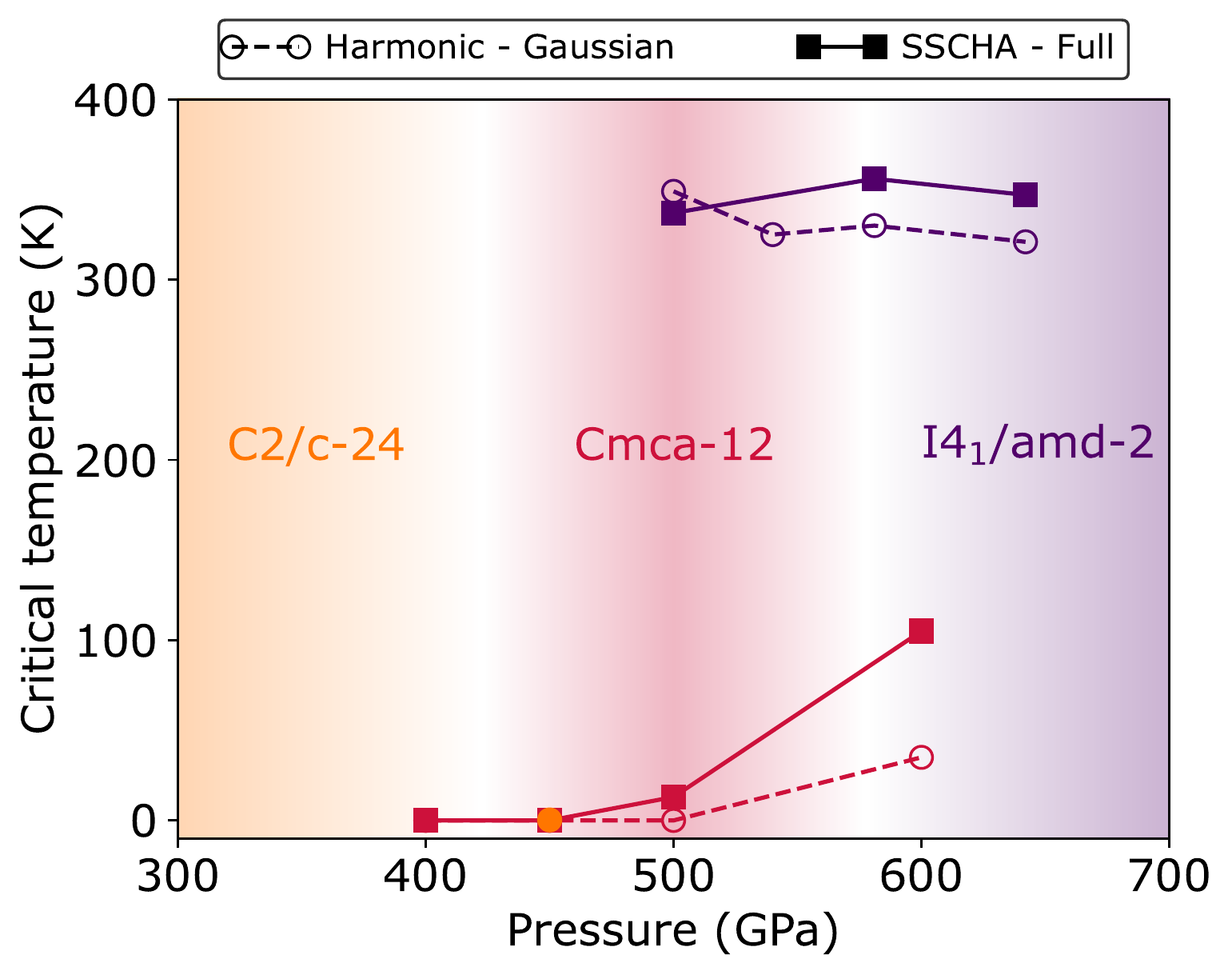}
			\caption{Calculated superconducting transition temperature in solid hydrogen in different phases and pressures within the SSCHA using the full phonon spectral functions (full symbols and solid lines) and the harmonic approximation using the Gaussian method (empty symbols and dashed lines). Shaded regions represent the phase diagram of solid hydrogen from Ref.~\cite{MonacelliPD}. Line colors denote for which phase calculations were performed (red for molecular phase VI and purple for atomic phase).}
			\label{fig5}
		\end{center}
	\end{figure}

	Solving isotropic Migdal-Eliashberg equations with the $\alpha^2F(\omega)$ obtained considering the full spectral function~\cite{AllenMitrovic, AnisoME}, we can estimate the impact of anharmonicity on the superconducting transition temperature (see  Fig.~\ref{fig5}). As mentioned above, the $C2/c-24$ phase of solid hydrogen does not exhibit superconducting behavior in the pressure range of interest. In the molecular phase VI the transition temperature is mostly linear with pressure and correlates well with the value of the DOS at the Fermi level. Because of this, the SSCHA structures consistently show higher transition temperatures than the classical harmonic ones. The difference in T$_{\text{C}}$ between these two methods increases with pressure, again due to the stronger dependence of the electronic DOS on the pressure in the SSCHA structures (see Fig.~\ref{fig1}), as well as due to the increased electron-phonon coupling due to the anharmonic softening of the phonon modes. 
	
	The estimate of the superconducting transition temperature obtained utilizing full phonon spectral function in all cases is larger than the one obtained using auxiliary SSCHA force constants and Gaussian approximation by about 30 K. On the other hand, Gaussian approximation coupled with the phonons from the Hessian of the total free energy gives a larger critical temperature than the full phonon spectral function calculation (at most 15 K). Since Hessian calculations only incorporate the softening of the phonon modes, the conclusion is that the softening of phonon modes increases the critical temperature while the broadening of the phonon spectral lines reduces it. Considering that $\alpha ^2F(\omega)$ is intimately related to the electron self-energy~\cite{AllenMitrovic} we can assume that the phonon spectral functions will have an influence on other material properties that strongly depend on the electron self-energy such as electrical conductivity, Seebeck coefficient, band gap renormalization, etc. We would like to highlight that at this moment, the effects of the finitely lived phonon quasiparticles are not accounted for in any first-principles calculations, while our results show they might have a large effect.
	
	Considering the critical dependence of  T$_{\text{C}}$ on the DOS at the Fermi level and that local exchange-correlation functionals tend to overestimate it~\cite{DoganC2c, DoganH, IonH, supH1, supH2, supH3}, we perform DFT calculations for the quantum SSCHA structures of phase VI using the B3LYP hybrid functional~\cite{b3lyp} (see Supplementary Material~\cite{supp_mat}). 
	Since the critical temperature correlates linearly with the electronic density of states in the $Cmca-12$ phase, we can estimate the superconducting transition temperature using the DOS from the better B3LYP calculation. With this procedure, we predict that superconductivity will emerge in solid hydrogen in the $Cmca-12$ phase between 450 and 500 GPa. This result is consistent with a recent experiment~\cite{H360} which failed to observe superconductivity at 440 GPa in what was identified as a molecular phase VI~\cite{Monacellimetal}.
	
	In the atomic tetragonal phase, the critical temperature is mostly constant with pressure. In this phase, T$_\mathrm{C}$ is mostly decorrelated with the value of the electronic DOS at the Fermi level because the structures are far away from the metal-insulator phase transition~\cite{MonacelliPD} and, despite quantum and anharmonic effects enhance the DOS as well, its relative increase is small compared to the molecular case. Accounting for the full phonon spectral function in the calculation of $\alpha^2F(\omega)$ increases the estimate of the critical temperature by 20 K compared to the case using the Gaussian approximation and SSCHA auxiliary force constants (see Supp. Material~\cite{supp_mat}). This increase is much larger than the one induced by the SSCHA structure renormalization (less than 5 K away from the structural instabilities, see Supp. Material). This highlights the important role that anharmonicity plays in the superconductivity of high-pressure hydrogen also in the atomic phase, contrary to the previous calculations that only estimated its effect within the Gaussian approximation of the spectral function \cite{IonH}.

	In conclusion, our first-principles calculations considering ionic quantum effects and anharmonicity show that superconductivity will emerge in solid hydrogen in molecular phase VI, between  450 and 500 GPa, and T$_{\mathrm{C}}$ will rapidly soar with pressure. We expect a jump of T$_{\text{C}}$ to approximately 350 K at the transition to the atomic phase. Quantum anharmonic effects have a huge impact on the structural, vibrational, and superconducting properties of both molecular and atomic phases by, for instance, increasing the H-H bonds and making the phonon spectral functions extremely broad and anomalous. We show that considering the full phonon spectral function in the calculation of $\alpha ^2F(\omega)$ enhances the predicted critical temperature by 25 K in the atomic phase and 30 K in the molecular phase VI.
	
	\noindent
	\textbf{Methods.}\\
	Density functional theory (DFT) and density functional perturbation theory (DFPT)~\cite{DFPT} calculations were performed using Quantum Espresso software~\cite{QE1, QE2}, implementing the generalized gradient approximation (GGA) with the BLYP parameterization~\cite{BLYP} for the exchange-correlation functional. In the case of the primitive unit cell calculations, we used a Monkhorst-Pack grid for sampling electronic states with densities of 48$\times$48$\times$48 for the atomic phase, 12$\times$12$\times$12 for the molecular phase VI, and 12$\times$6$\times$12 for molecular phase III. The electronic wave functions were represented in a plane wave expansion using an 80 Ry energy cutoff (320 Ry cutoff for the electronic density). To describe hydrogen ions we used a norm-conserving pseudo-potential with no pseudized electrons generated by the Pseudo Dojo library~\cite{Pseudodojo} and the ONCVPSP software~\cite{ONCVP}. Considering that we are investigating metallic/semimetallic phases we used a Marzari–Vanderbilt smearing of 0.03 Ry~\cite{MarzariVanderbilt} for Brillouin zone integrations.
	
	To get the structural and vibrational properties of solid hydrogen we used the stochastic self-consistent harmonic approximation (SSCHA). The SSCHA method~\cite{SSCHA1,SSCHA2, SSCHA3, SSCHA4} allows us to minimize the total free energy of the system, which includes the quantum zero-point motion and anharmonicity, with respect to two variational parameters that define the ionic wave function: the centroid positions and the auxiliary force constants. The centroids are the average positions of the atoms (the means of the Gaussians that approximate the ionic wave functions). The auxiliary force constants are related to the standard deviation of the Gaussians. Eigenvalues of the dynamical matrices constructed from these auxiliary force constants can be regarded as better estimates of the true phonon frequencies than the simple harmonic force constants since they have been renormalized by anharmonicity. More precisely, in perturbation theory language these force constants include contributions to the first order in perturbative expansion from all of the anharmonic terms in the expansion of the BOES. These corrections are purely real and only shift the phonon frequency. The centroids and SSCHA auxiliary second-order force constants are obtained at the end of the minimization of the total free energy. Additionally, on top of renormalizing the second-order force constants, SSCHA renormalizes the anharmonic force constants as well in a similar manner. 
	
	From here, we can go a step further and include some terms of the higher orders in the perturbation theory that stem from third and fourth-order anharmonic force constants (renormalized by anharmonicity as explained above) which are consistent with SSCHA~\cite{SSCHA4, TDSSCHA, TDSSCHALihm, Siciliano}. The phonon Green's function ($G _{\mu\mu'}(\mathbf{q}, \omega)$) in this case can be expressed as:
	\begin{align*}
	G _{\mu\mu'}(\mathbf{q}, \Omega) = \left[\Omega ^2\delta_{\mu\mu'} - \stackrel{(2)}{D} _{\mu\mu'}(\mathbf{q}) - \Pi _{\mu\mu'}(\mathbf{q} ,\Omega)\right]^{-1}.
	\end{align*}
	Here $\stackrel{(2)}{D} _{\mu\mu'}(\mathbf{q})$ is the dynamical matrix constructed from the SSCHA auxiliary force constants and $\Pi _{\mu\mu'}(\mathbf{q},\omega)$ is phonon self-energy that depends on the SSCHA anharmonic force constants ($ \stackrel{(3)}{\boldsymbol{\mathcal{D}}}(\mathbf{q}), \stackrel{(4)}{\boldsymbol{\mathcal{D}}}(\mathbf{q})$):
	\begin{align*}
	\boldsymbol{\Pi}(\mathbf{q}, \Omega) = \stackrel{(3)}{\boldsymbol{\mathcal{D}}}(\mathbf{q}):\boldsymbol{\Lambda}(\mathbf{q},\Omega):\left[ \boldsymbol{\mathbf{1}} -  \stackrel{(4)}{\boldsymbol{\mathcal{D}}}(\mathbf{q}):\boldsymbol{\Lambda}(\mathbf{q}, \Omega)\right]^{-1}:\stackrel{(3)}{\boldsymbol{\mathcal{D}}}(\mathbf{q}). 
	\end{align*}
	The double-dot product $\mathbf{X}:\mathbf{Y}$ indicates the contraction of the last two indices of $\mathbf{X}$ with the first two indices of $\mathbf{Y}$. If we denote the eigenvalues of the SSCHA auxiliary dynamical matrices as $\omega _{\mu}(\mathbf{q})$ and associated Bose-Einstein factors as $n_{\mu}(\mathbf{q})$, the above $\boldsymbol{\Lambda}(\mathbf{q}, \Omega)$ is given as:
	\begin{widetext}
		\begin{equation}
		\Lambda ^{\mu\mu'}(\mathbf{q}, \Omega) = \frac{1}{4\omega _{\mu}(\mathbf{q})\omega _{\mu'}(\mathbf{q})}\left[\frac{\left(\omega _{\mu}(\mathbf{q}) - \omega _{\mu'}(\mathbf{q})\right)\left(n_{\mu}(\mathbf{q}) - n_{\mu'}(\mathbf{q})\right)}{(\omega _{\mu}(\mathbf{q}) - \omega _{\mu'}(\mathbf{q}))^2 - \Omega ^2 + i\epsilon} - \frac{(\omega _{\mu}(\mathbf{q}) + \omega _{\mu'}(\mathbf{q}))(1 + n_{\mu}(\mathbf{q}) + n_{\mu'}(\mathbf{q}))}{(\omega _{\mu}(\mathbf{q}) + \omega _{\mu'}(\mathbf{q}))^2-\Omega ^2 + i\epsilon}\right]. 
		\label{eq:lambda}
		\end{equation}
	\end{widetext}
	$\Pi _{\mu\mu'}(\mathbf{q},\omega)$ is not purely real and describes the realistic broadening of the phonon spectral functions. However, in the static limit ($\Omega\rightarrow 0$), the contributions from these terms are again only real and can be included to further renormalize the SSCHA second-order auxiliary force constants. Force constants obtained in this manner are Hessians of the total free energy, $G _{\mu\mu'}(\mathbf{q}, 0)$. If any of the eigenvalues of the Hessian of the total free energy is imaginary, the structure is unstable. These force constants can alternatively be used to describe the vibrational properties of the material. In the static limit, for the calculation of the Hessian of total free energy, we include the contributions of both the third and fourth-order SSCHA anharmonic force constants.
	
	However, a physically more relevant representation of the vibrational properties of materials comes from the phonon spectral functions obtained in the dynamical dressed-bubble approximation, using auxiliary force constants and third-order force constants from SSCHA as described in Refs.~\cite{SSCHA4, TDSSCHA, TDSSCHALihm, Siciliano}:
	\begin{align*}
	\sigma _{\mu\mu'}(\mathbf{q}, \Omega) = -\frac{\Omega}{\pi}G _{\mu\mu'}(\mathbf{q}, \Omega).
	\end{align*}
	The anharmonicity in general leads to the mixing of the phonon modes and the matrices of phonon spectral functions at different values of the frequency (energy) $\Omega$ do not commute. Usually, this is disregarded and only the diagonal part $\mu=\mu'$ of the phonon spectral function (in the space of eigenvectors that diagonalize auxiliary SSCHA force constants) is taken into account. This approximation is referred to as a "no-mode-mixing" approximation in this work. Alternatively, one can use the true phonon spectral function, including the off-diagonal terms in the phonon spectral functions, and that approach is termed "full" in this work (see Supp. Material for more information~\cite{supp_mat}).
	
	The sampling of atomic positions and forces was done on a 5$\times$5$\times$5 primitive cell repetition for the atomic phase, 2$\times$2$\times$2 for the molecular phase VI, and 2$\times$1$\times$2 for the molecular phase III. The number of configurations used for the stochastic sampling was 500 for the atomic phase, 600 for molecular phase VI, and 6000 for molecular phase III. To calculate third-order force constants needed to calculate the spectral functions we used a finer stochastic sampling of 3000 structures for the atomic phase and 20000 structures for phase VI. SSCHA calculations were performed at 0 K. For the calculation of the phonon spectral functions we used the dynamical bubble term in the phonon self-energy expansion. In the static limit, the peaks of the phonon spectral function coincide with the frequencies obtained from the free energy Hessian. For the Hessian calculations, in the molecular phase, we used the static bubble term from the third-order anharmonicity and fourth-order anharmonicity double bubble term, and for the atomic phase, we used only the third-order static bubble term. The SSCHA auxiliary force constants already include the effects of so-called tadpole and loop terms, as well as higher orders of anharmonicity.
	
	Finally, we performed a convergence study of the electron-phonon coupling constant and the critical temperature with respect to the $\mathbf{q}$ point grid in DFPT calculations. We have found that reasonably converged results were obtained with a 12$\times$12$\times$12 $\mathbf{q}$ point grid for the atomic phase, 8$\times$8$\times$8 for phase VI, and 8$\times$4$\times$8 for phase III. The calculated electron-phonon coupling constants from DFPT were projected onto SSCHA phonon modes~\cite{IonPdH}. $\mathbf{k}$-point grids for the non-self-consistent calculations for the electron-phonon coupling were done on 100$\times$100$\times$100 grids for the atomic phase, 44$\times$44$\times$44 for phase VI, and 44$\times$22$\times$44 for phase III with Gaussian smearing of 0.012 Ry for the energy conservation Dirac deltas. Finally, to calculate superconducting transition temperatures we used the isotropic approximation of the Migdal-Eliashberg equations in the constant density of states approximation~\cite{AllenMitrovic}. We use $\mu ^* = 0.16$ for the Coulomb pseudopotential and a cutoff for the Matsubara frequencies of 10 times the highest phonon frequency~\cite{AllenMitrovic}. We have checked that this approximated approach to solve Migdal-Eliashberg (ME) equations yields accurate results despite the use of the $\mu ^*$ parameter. For example, in LaH$_{10}$, where superconductivity is dominated by the hydrogen sublattice, this approach only yields an overestimation of T$_{\mathrm{C}}$ of a 7\% with respect to anisotropic ME equations and the use of the random phase approximation to calculate the Coulomb repulsion (to avoid the use of the simple $\mu ^*$ parameter)~\cite{IonLaH,Pellegrini2022Eliashberg}.
	
	\noindent
	\textbf{Data and code availability.}\\
	Both Quantum Espresso and SSCHA are free software codes freely available from the following websites: \href{https://www.quantum-espresso.org}{www.quantum-espresso.org} and \href{http://sscha.eu}{sscha.eu}. All the data supporting the presented results are available from the corresponding author upon request.
	
	\bibliography{arxiv_version}

\begin{thebibliography}{61}
\bibitem [{Pre()}]{PressureScale}%
  \BibitemOpen
  \href@noop {} {}\bibinfo {note} {Experimental estimation of the applied
  pressure depends on different pressure scales used in the community. For this
  reason, the actual pressure exerted on the hydrogen could be
  lower.}\BibitemShut {Stop}%
\end{thebibliography}
	
	\noindent
	\textbf{Acknowledgements.}\\
	This work is supported by the European Research Council (ERC) under the European Unions Horizon 2020 research and innovation program (grant agreement No. 802533), the Department of Education, Universities and Research of the Eusko Jaurlaritza and the University of the Basque Country UPV/EHU (Grant No. IT1527-22), and the Spanish Ministerio de Ciencia e Innovacion (Grant No. PID2022-142861NA-I00). L.M. acknowledges the European Union MSCA-IF fellowship for funding the project THERMOH. We acknowledge PRACE for awarding us access to the EuroHPC supercomputer LUMI located in CSC’s data center in Kajaani, Finland through EuroHPC Joint Undertaking (EHPC-REG-2022R03-090).
	
	\noindent
	\textbf{Author contributions.}\\
	{\DJ}.D. and I.E. conceived the research plan. {\DJ}.D. performed the first principles calculations. {\DJ}.D. and I.E. wrote the article with input from all the authors. All authors discussed and interpreted the results.
	
	\noindent
	\textbf{Competing interests.}\\
	The authors declare no competing interests.
	
	\onecolumngrid
	\renewcommand{\figurename}{Supplementary Figure}
	\renewcommand{\tablename}{Supplementary Table}
	\setcounter{figure}{0}
	\newpage
	\newpage
	\newpage

	\section{Supplementary Information for \\ ``Large impact of phonon lineshapes on the superconductivity of solid hydrogen''}
	\subsection{Comparison of harmonic and SSCHA phonon band structures}
	
\begin{figure*}[!t]
		\begin{center}
			\includegraphics[width=0.95\linewidth]{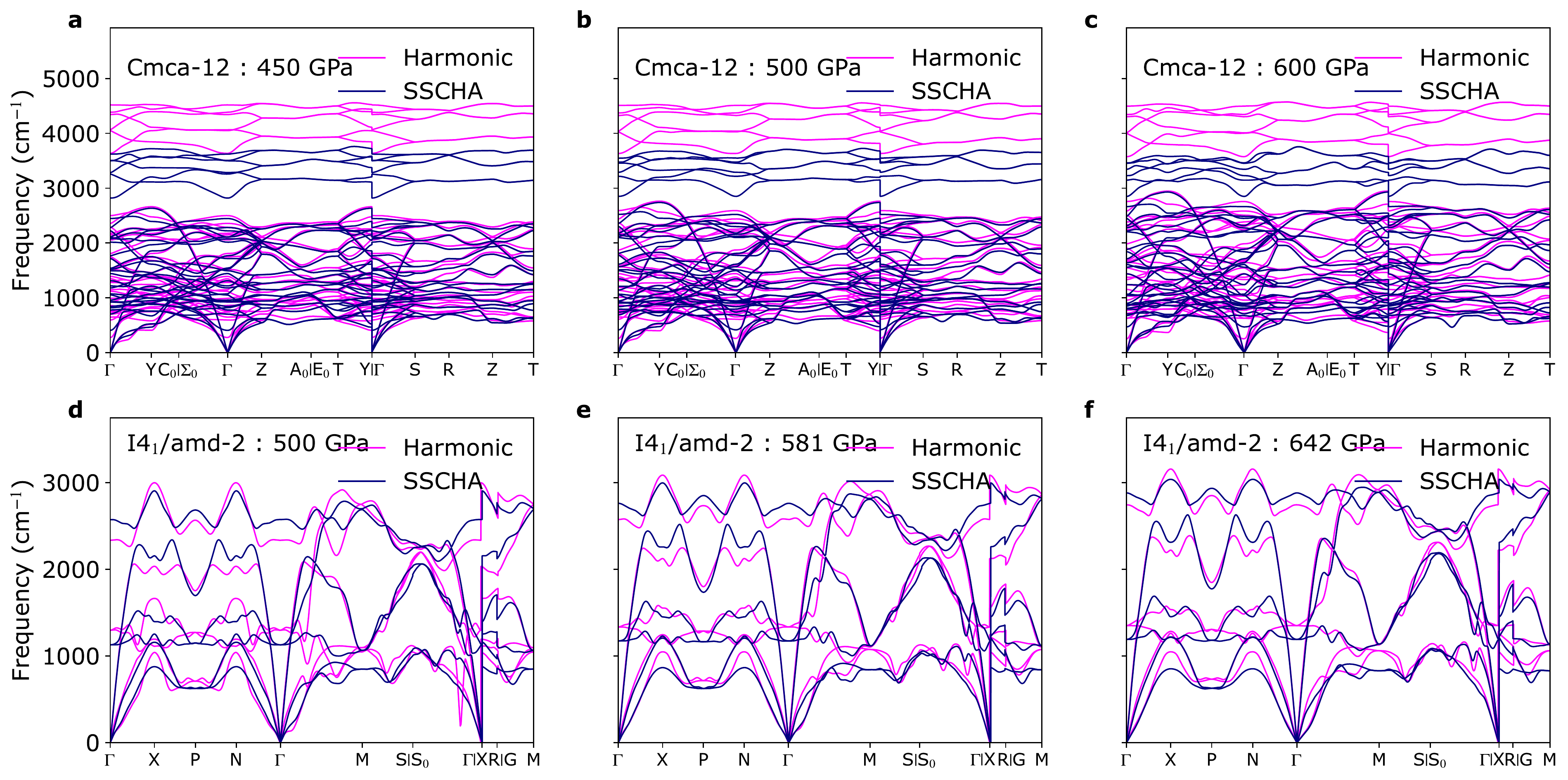}
			\caption{Phonon band structures for harmonic and SSCHA structures of solid hydrogen in molecular Cmca-12 phase VI at (a) 450 GPa, (b) 500 GPa, (c) 600 GPa, and atomic I4$_1$/amd-2 tetragonal phase at (d) 500 GPa, (e) 581 GPa, and (f) 642 GPa. SSCHA phonon spectra are calculated from the auxiliary force constants.}
			\label{sfig1}
		\end{center}
	\end{figure*}

	In Supp. Fig.~\ref{sfig1} we show the comparison between harmonic and SSCHA phonon band structures for relevant systems (systems that have non-zero superconducting critical temperatures). In the harmonic case, we calculated the phonons using DFPT for the structures that minimize the Born-Oppenheimer energy. In the SSCHA case, we are showing the eigenvalues of the SSCHA auxiliary dynamical matrices for the structures that minimize the total free energy in the self-consistent harmonic approximation. 
	
	As we note in the main text, we can see a large softening of the phonon frequencies in the case of the SSCHA structures. This is particularly prominent for optical modes in molecular Cmca-12 phase VI. Another prominent difference is that SSCHA cures the incipient instability in atomic hydrogen at 500 GPa on the S$_0\rightarrow\Gamma$ line, which leads to the increase of the electron-phonon coupling strength in the harmonic case at this pressure.
	
	\newpage
	\subsection{Dependence of the critical temperature on ground state properties}
	
\begin{figure}[h]
		\begin{center}
			\includegraphics[width=0.95\linewidth]{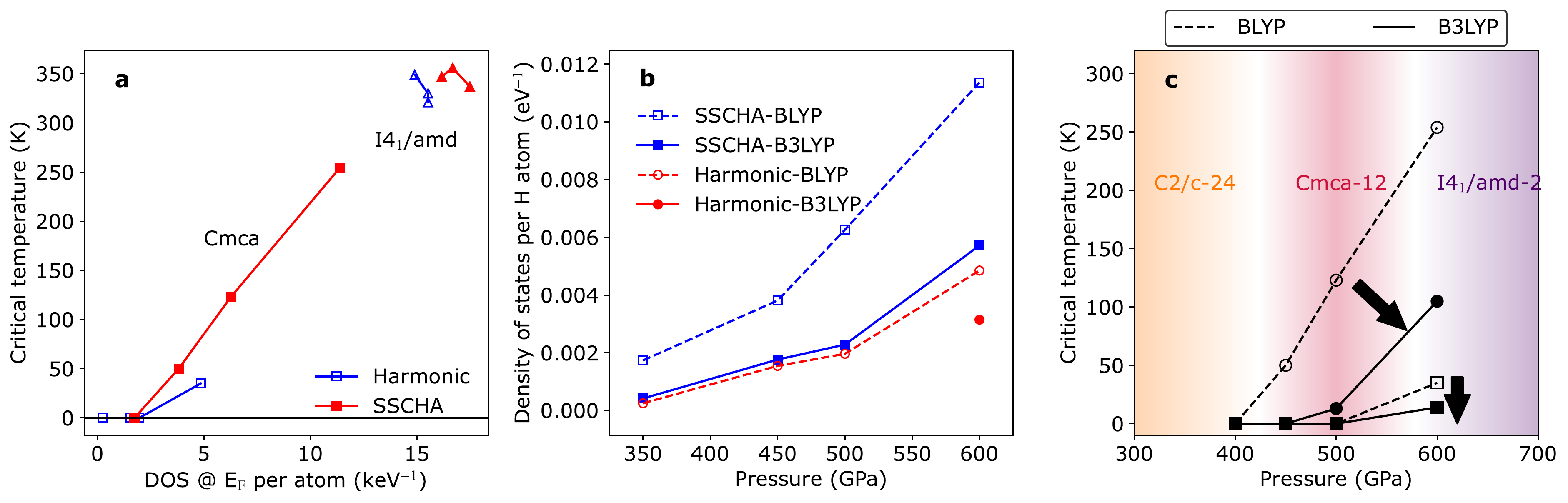}
			\caption{a) Critical temperature as a function of the value of the electronic density of states at the Fermi level in molecular phase VI and atomic phase of solid hydrogen. $\alpha ^2F$ needed for the estimation of T$_{\mathrm{C}}$ was obtained in the Gaussian approximation for the harmonic case and using full phonon spectral function in the SSCHA case. (b) Density of states at the Fermi level of solid hydrogen in Cmca-12 phase (molecular phase VI) calculated with BLYP and B3LYP approximations for the exchange-correlation functional. (c) The changes in the critical temperature in the molecular phase VI (Cmca-12) due to different exchange-correlation functionals. Circles represent results obtained for SSCHA structures with fully anharmonic Eliashberg spectral functions, while squares are for harmonic structures and Eliashberg spectral functions in Gaussian approximation.}
			\label{sfig3}
		\end{center}
	\end{figure}

	In Supp. Fig.~\ref{sfig3} (a) we show the correlation between the superconducting critical temperature T$_\mathrm{C}$ and the electronic density of states at the Fermi level in these compounds. As we can see, in molecular phase VI there is a linear correlation between T$_\mathrm{C}$ and the electronic density of states. The critical temperature in the figure was calculated using isotropic Migdal-Eliashberg equations with $\alpha ^2F$ calculated with full phonon spectral functions in the SSCHA case and Gaussian approximation in the harmonic case. 
	
	In the atomic phase, it looks like there is a negative correlation between the critical temperature and the electronic density of states. The decrease of the critical temperature in these cases is probably due to the stiffening of the phonon modes with the increased pressure and probably is not connected to the changes in the density of states. This increase in phonon frequencies decreases the electron-phonon coupling strength, which, in turn, reduces T$_{\mathrm{C}}$. The increase of the electronic density of states does not influence the critical temperature as strongly in this phase since it is far away from the metal-insulator phase transition.
	
	To estimate the error of the calculated density of states using BLYP, we performed  B3LYP calculations of the electronic structure on the SSCHA structures that have non-zero critical temperatures. The results are shown and compared to DFT in Supp. Fig.~\ref{sfig3} (b). As we can see the results with B3LYP drastically reduce the calculated density of states at the Fermi level. Accidentally, B3LYP values agree quite well with BLYP values calculated for harmonic structures. We have also calculated the electronic density of states with B3LYP hybrid functional for 600 GPa harmonic structure (the only one that shows superconductivity in Cmca-12 phase for harmonic structure and Gaussian approximation) and found that it decreases the DOS at the Fermi level significantly. 
	
	If we assume that the critical temperature is directly proportional to the density of states at the Fermi level we can  estimate that the onset of the superconductivity in this phase happens between 450 and 500 GPa. This is the result reported in the main text.
	
\begin{figure}
		\centering
		\includegraphics[width=0.9\textwidth]{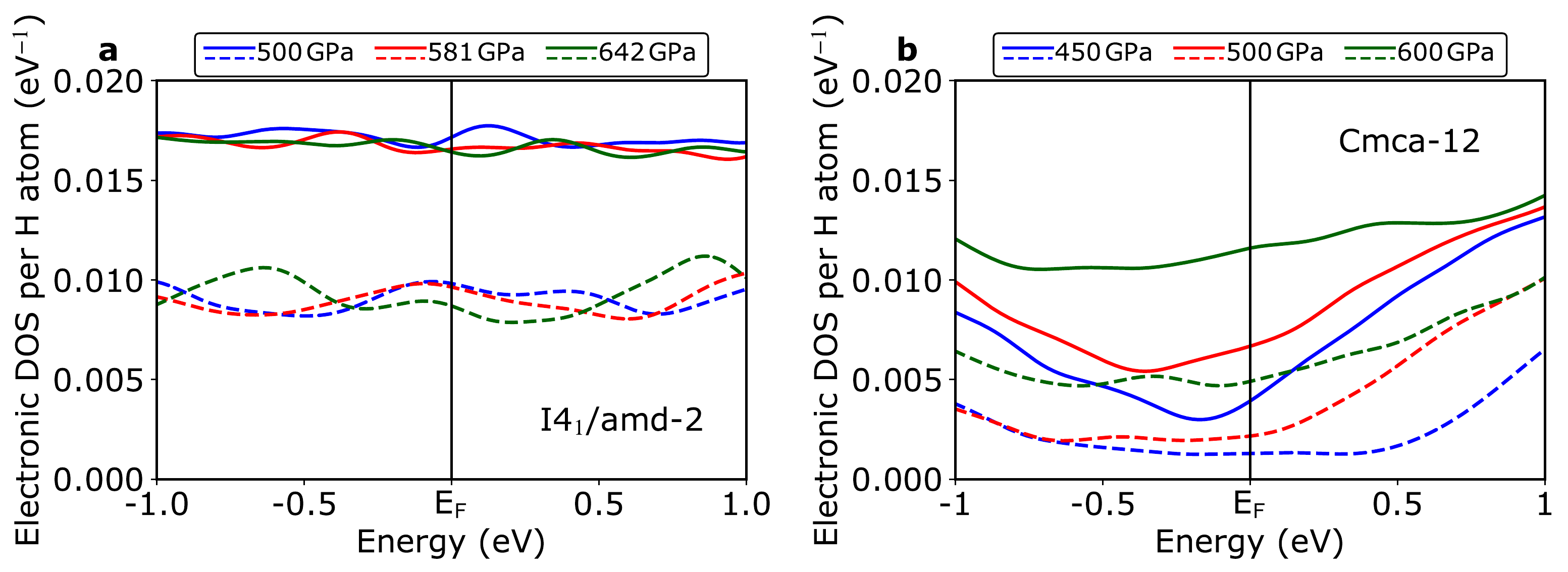}
		\caption{Electronic density of states calculated with B3LYP (dashed lines) and BLYP (full lines) exchange-correlation functionals for SSCHA structure of solid hydrogen in \textbf{a} atomic and \textbf{b} molecular VI phase.}
		\label{edos}
	\end{figure}

	The comparison between BLYP and B3LYP electronic density of states is given in Supp. Fig.~\ref{edos}. As we have already mentioned, B3LYP in all cases gives a significantly lower density of states at the Fermi level. The density of states in the atomic phase and highest pressure in the molecular phase has a pretty constant profile in the energy window of interest ($\pm 2\omega_{max}$), justifying the use of frozen Fermi level approximation for the calculation of critical temperatures. At lower pressures in molecular phase VI there is some change in the electronic density of states in this energy window, but it is not expected to have a drastic effect on the estimated critical temperature.
	
	\newpage
	\subsection{Phonon spectral functions in solid hydrogen}
	
\begin{figure}[h]
		\begin{center}
			\includegraphics[width=0.85\linewidth]{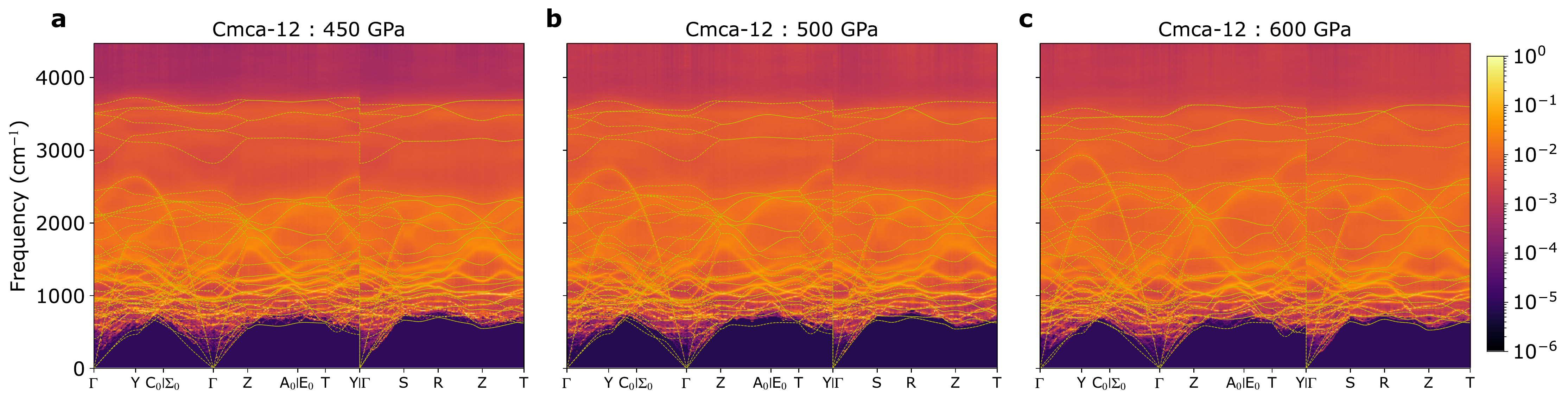}
			\includegraphics[width=0.85\linewidth]{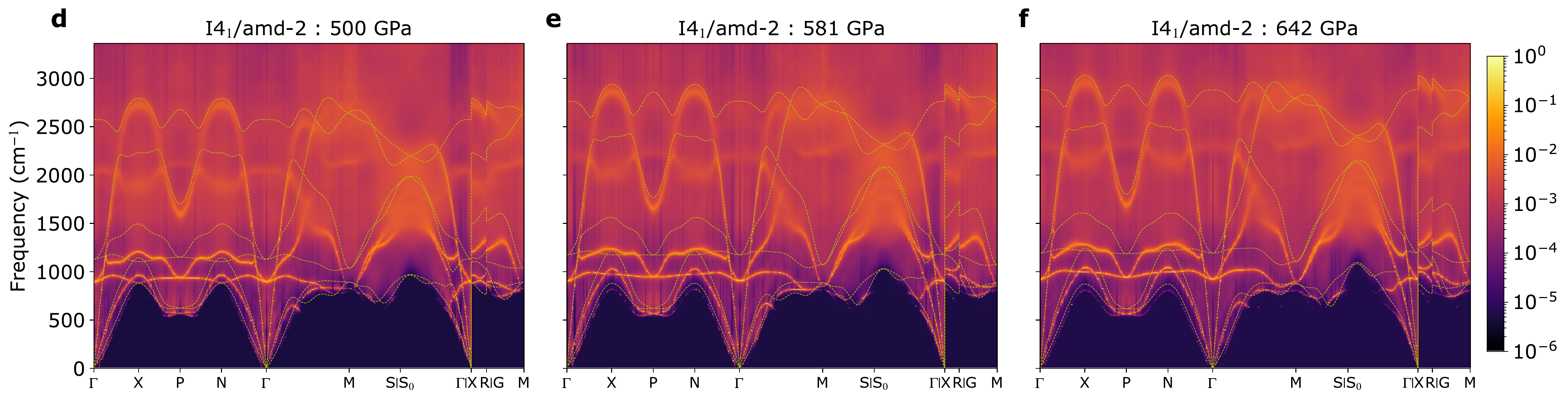}
			\caption{Phonon spectral function $\sigma(\mathbf{q},\omega)=\sum_\mu \sigma_{\mu}(\mathbf{q},\omega)$ calculated in the no-mode mixing approximation in arbitrary units of solid hydrogen in molecular phase VI (Cmca-12)  at (a) 450 GPa, (b) 500 GPa, (c) 600 GPa, and tetragonal I4$_1$/amd-2 atomic phase at (d) 500 GPa, (e) 581 GPa, and (g) 642 GPa. Dashed yellow lines represent eigenvalues of the SSCHA auxiliary dynamical matrices.}
			\label{specfunc}
		\end{center}
	\end{figure}

	In Supp. Fig.~\ref{specfunc} we are showing the phonon lineshapes of solid hydrogen at different pressures and phases along some high symmetry lines in the reciprocal space.  Anharmonicity further softens the phonon frequencies for most of the modes, most prominently for the optical phonon modes. There is an obvious closing of the gap between optical modes in the molecular phase of hydrogen. 
	
\begin{figure*}[h]
		\begin{center}
			\includegraphics[width=0.95\textwidth]{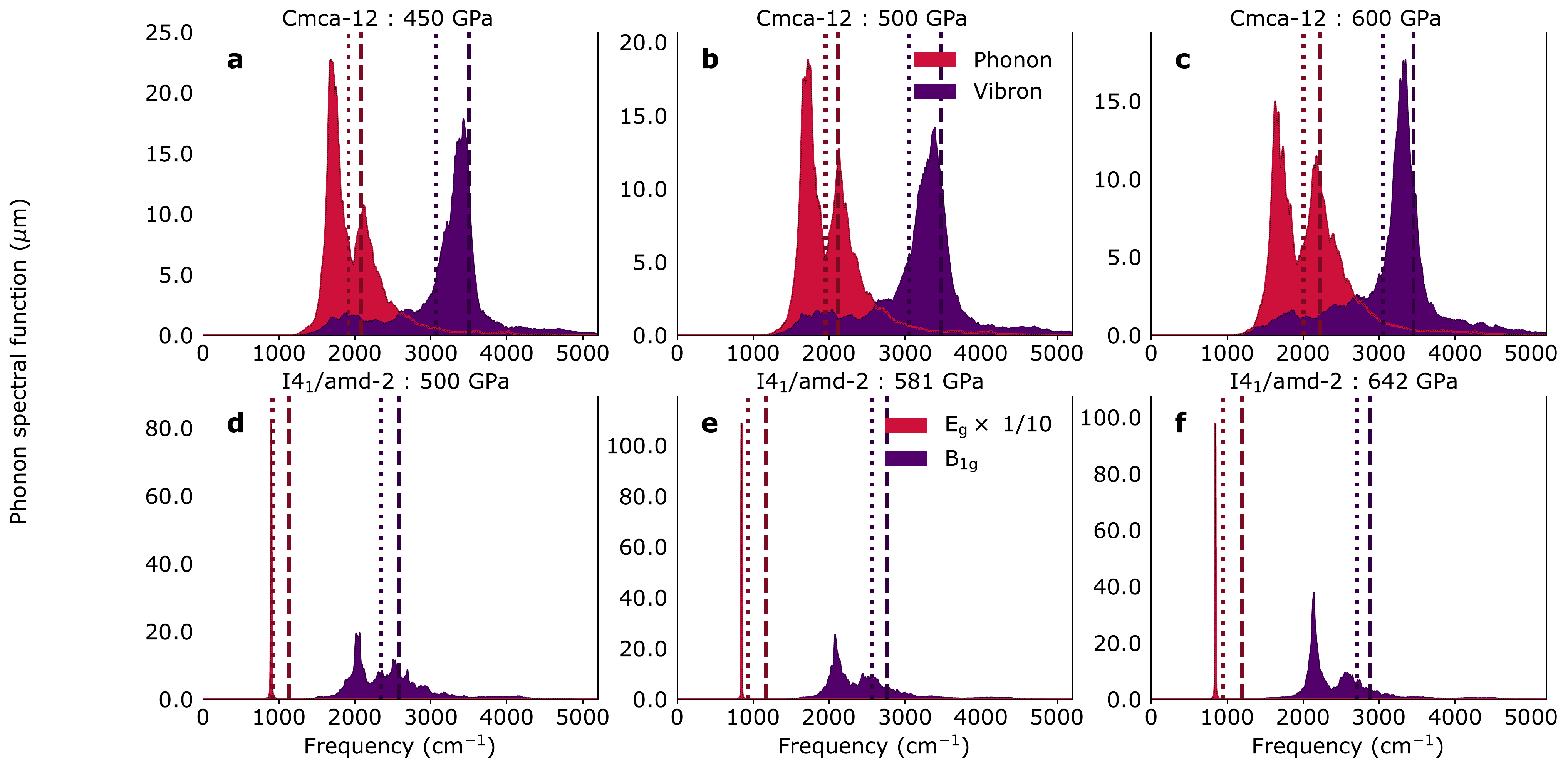}
			\caption{Phonon spectral functions in the no mode mixing approximation in modes basis, $\sigma_{\mu}(\mathbf{q},\omega)$, of two representative optical phonon modes at $\Gamma$ of solid hydrogen in molecular Cmca-12 phase VI at (a) 450 GPa, (b) 500 GPa, (c) 600 GPa, and atomic tetragonal I4$_1$/amd-2 phase at (d) 500 GPa, (e) 581 GPa, and (f) 642 GPa. In figures (d), (e), and (f) we scaled the values of the E$_\mathrm{g}$ mode in order to make the figures clearer. Thick dashed vertical lines represent the corresponding frequencies obtained from the auxiliary SSCHA force constants.}
			\label{specfunc2}
		\end{center}
	\end{figure*}

	In Supp. Fig.~\ref{specfunc2} we see the phonon mode spectral functions of solid hydrogen at $\Gamma$ for representative modes. All phases at all pressures show large phonon lineshifts and linewidths.
	
	\newpage
	\newpage
	sub\section{Comparison between anharmonic and gaussian Eliashberg spectral functions for SSCHA structures} 
	
	The solution to the isotropic Migdal-Eliashberg equations gives us the value of the superconducting gap as a function of temperature. The temperature where this value drops to zero we call the critical temperature. To solve isotropic Migdal-Eliashberg equations one only needs the Eliashberg spectral  function. 
	
	The general definition of the Eliashberg spectral function ($\alpha ^2F (\omega)$) is~\cite{AllenMitrovic}:
	\begin{align}
		\alpha ^2F _{nn'} (\mathbf{k},\mathbf{q},\omega) = \frac{N_{\mathrm{F}}}{N_{\mathbf{k}}N_{\mathbf{q}}}\sum _{a,b}  d^{a}_{n\mathbf{k},n'\mathbf{k}+\mathbf{q}}d^{b}_{n'\mathbf{k}+\mathbf{q},n\mathbf{k}}B_{ab}(\mathbf{q},\omega).
		\label{a2f_gen}
	\end{align}
	Here $a,b$ compactly label atoms in the primitive cell and Cartesian directions, $d^{a}_{n\mathbf{k},n'\mathbf{k}+\mathbf{q}}$ is the deformation potential $d^{a}_{n\mathbf{k}, n'\mathbf{k}+\mathbf{q}} = \langle n\mathbf{k} \vert \frac{\delta V}{\delta u ^{a}(\mathbf{q})}\vert n'\mathbf{k} + \mathbf{q} \rangle$, with $\vert n\mathbf{k} \rangle$ the Kohn-Sham state of band $n$ and wave number $\mathbf{k}$, and $B_{ab}(\mathbf{q},\omega)$ is defined as:
	\begin{align*}
		B_{ab}(\mathbf{q},\omega) = -\frac{1}{\pi} \mathrm{Im}\mathcal{D}_{ab}(\mathbf{q},\omega),
	\end{align*}
	where $\mathcal{D}_{ab}(\mathbf{q},\omega)$ is the Fourier transform of the phonon Green's function: $\mathcal{D}_{ab}(\mathbf{q},t) = - \langle Tu_a (\mathbf{q},t)u_b^{*} (\mathbf{q},0)\rangle$. In order to get the isotropic Eliashberg spectral function from Eq.~\ref{a2f_gen}, we average $\alpha ^2F _{nn'} (\mathbf{k},\mathbf{q},\omega)$ over the Fermi surface. Once we do that, we obtain:
	\begin{align}
		\alpha ^2F(\omega) = \frac{1}{N_{\mathbf{q}}}\sum _{a,b,\mathbf{q}} \Delta ^{ab}(\mathbf{q})B_{ab}(\mathbf{q},\omega).
		\label{eq:a2F}
	\end{align}
	Here $\Delta ^{ab}(\mathbf{q})$ is the shorthand notation for the deformation potential averaged over Fermi surface:
	\begin{align*}
		\Delta ^{ab}(\mathbf{q}) = \frac{1}{N_{\mathrm{F}}N_{\mathbf{k}}}\sum_{n,n',\mathbf{k}}d^{a}_{n\mathbf{k},n'\mathbf{k}+\mathbf{q}}d^{b}_{n'\mathbf{k}+\mathbf{q},n\mathbf{k}}\delta(\epsilon _{n\mathbf{k}} - \epsilon _{\mathrm{F}})\delta(\epsilon _{n'\mathbf{k}+\mathbf{q}} - \epsilon _{\mathrm{F}}).
	\end{align*}
	
	In the SSCHA code we use a slightly different definition of the phonon Green's function compared to the definition used here. We define the Green's function with respect to the displacement of the atom scaled by the square root of the atom mass. Additionally, we define the phonon spectral function $\sigma _{ab}(\mathbf{q},\omega)$ as:
	\begin{align*}
		\sigma _{ab}(\mathbf{q},\omega) = -\frac{\omega}{\pi}\mathrm{Im}G_{ab}(\mathbf{q},\omega) =\frac{\omega}{\pi}\mathrm{Im}\sqrt{m_{a}m_{b}}\langle Tu_a u_b^*\rangle (\mathbf{q}, \omega) = \omega\sqrt{m_{a}m_{b}}B_{ab}(\mathbf{q},\omega) .
	\end{align*}
	This result is given in the atomic Cartesian basis, but we can cast this result in the no mode mixing approximation. In this approximation atomic displacements are projected onto phonon modes ($\sigma _{ab}(\mathbf{q},\omega) = \sum _{\mu}e_{\mu} ^{a}(\mathbf{q})e _{\mu} ^{b*}(\mathbf{q})\sigma_{\mu}(\mathbf{q},\omega)$). This no mode mixing approximation is usually very good in describing phonon spectral functions. 
	
	Usually, the mode projected phonon spectral function $\sigma_{\mu}(\mathbf{q})$ is approximated with:
	\begin{align}
		\sigma^h_{\mu}(\mathbf{q},\omega) = \frac{1}{2}\left[\delta (\omega + \omega^h _{\mu}(\mathbf{q})) + \delta (\omega - \omega^h _{\mu}(\mathbf{q}))\right].
		\label{eq_dirac}
	\end{align}
	Then Dirac delta $\delta (\omega - \omega^h _{\mu}(\mathbf{q}))$ is approximated with a Gaussian centered at the value of harmonic phonon frequency $\omega^h _{\mu}(\mathbf{q})$ and the fixed width which is the same for each phonon mode (in our calculation we took this smearing parameter to be 10 cm$^{-1}$) .
	
	SSCHA however, allows us to calculate the third-order interatomic force constants, in addition to the crystal structure and second order force constants renormalized by anharmonicity. Using these third-order force constants we can explicitly calculate the phonon mode spectral function $\sigma_{\mu}(\mathbf{q},\omega)$ (in the diagonal dynamical bubble approximation). It is defined as~\cite{SSCHA1, SSCHA4}:
	\begin{align}
		\sigma_{\mu} (\mathbf{q},\omega) = \frac{1}{2\pi}\left[ \frac{-\mathrm{Im}\mathcal{Z}_{\mu}(\mathbf{q},\omega)}{[\omega - \mathrm{Re}\mathcal{Z}_{\mu}(\mathbf{q},\omega)]^2 + [\mathrm{Im}\mathcal{Z}_{\mu}(\mathbf{q},\omega)]^2} + \frac{\mathrm{Im}\mathcal{Z}_{\mu}(\mathbf{q},\omega)}{[\omega + \mathrm{Re}\mathcal{Z}_{\mu}(\mathbf{q},\omega)]^2 + [\mathrm{Im}\mathcal{Z}_{\mu}(\mathbf{q},\omega)]^2}\right].
		\label{eq_full}
	\end{align}
	Here $\mathcal{Z}_{\mu}(\mathbf{q},\omega) = \sqrt{\omega_{\mu}^2(\mathbf{q}) + \Pi _{\mu}(\mathbf{q}, \omega)}$, where $\Pi _{\mu}(\mathbf{q}, \omega)$ is the phonon self-energy due to phonon-phonon interaction in the bubble approximation:
	\begin{align*}
		\Pi_{\mu}(\mathbf{q}, \omega) &= \frac{1}{N_\mathbf{k}}\sum _{\mathbf{k}_{1},\nu}\sum _{\mathbf{k}_{2},\rho}\sum _{\mathbf{G}}\delta _{\mathbf{G}, \mathbf{q} + \mathbf{k}_{1} + \mathbf{k}_{2}}\lvert D_{\mathbf{q}\mathbf{k}_{1}\mathbf{k}_{2}}^{\mu\nu\rho}\rvert^2\times \\
		&\times\frac{\hbar}{4\omega _{\nu}(\mathbf{k}_{1})\omega _{\rho}(\mathbf{k}_{2})}\left(\frac{(\omega _{\nu}(\mathbf{k}_{1})-\omega _{\rho}(\mathbf{k}_{2}))(n _{\nu}(\mathbf{k}_{1}) - n _{\rho}(\mathbf{k}_{2}))}{(\omega _{\nu}(\mathbf{k}_{1})-\omega _{\rho}(\mathbf{k}_{2}))^2 - \omega ^2} - \frac{(\omega _{\nu}(\mathbf{k}_{1})+\omega _{\rho}(\mathbf{k}_{2}))(n _{\nu}(\mathbf{k}_{1}) + n _{\rho}(\mathbf{k}_{2}) + 1)}{(\omega _{\nu}(\mathbf{k}_{1})+\omega _{\rho}(\mathbf{k}_{2}))^2 - \omega ^2}\right). \numberthis
		\label{eq_self_energy}
	\end{align*}
	Here $n _{\mu}(\mathbf{q})$ is the Bose-Einstein occupation factor for the phonon mode with frequency $\omega _{\mu}(\mathbf{q})$ and $D_{\mathbf{q}\mathbf{k}_{1}\mathbf{k}_{2}}^{\mu\nu\rho}$ is the Fourier transform of the third-order force constants (including the scaling with atom masses). An important note is that $\omega _{\mu}(\mathbf{q})$ are calculated from eigenvalues of the auxiliary SSCHA force constants. The same definition of the spectral function was used in Fig. 2 of the main text. As we can see this quantity is actually temperature dependent (through $n _{\mu}(\mathbf{q})$) and in principle should be recalculated at each temperature. Here, however, we only calculate it at 0 K ($n _{\mu}(\mathbf{q}) = 0$ always), and the only processes accounted for are the annihilations of two phonons (second term in the parenthesis). The non-zero temperature would only make changes for phonon modes with frequencies lower than $k_BT$. Since the relevant temperature scale for this study is up to 300 K (200 cm$^{-1}$), including temperature will not make any significant change for any of the phonons in our $\mathbf{q}$ point grid.
	
	We performed calculations for atomic hydrogen at 500 GPa using spectral functions calculated at 300 K and the results for the critical temperature did not change. Another important detail that we would like to stress is that the phonon spectral functions calculated here come purely from the phonon-phonon interaction. We justify this with the fact that for the temperature range of interest (around 100 K), phonon linewidths due to the phonon-phonon interaction are orders of magnitude larger than the phonon linewidths due to the electron-phonon interaction for most of the phonon modes. Since phonon-phonon self-energy increases with temperature and electron-phonon, in general, does not, if phonon-phonon interaction is a dominant contribution to phonon linewidths at 100 K, this conclusion will hold at even higher temperatures.
	
	\begin{figure*}[!t]
		\begin{center}
			\includegraphics[width=0.95\textwidth]{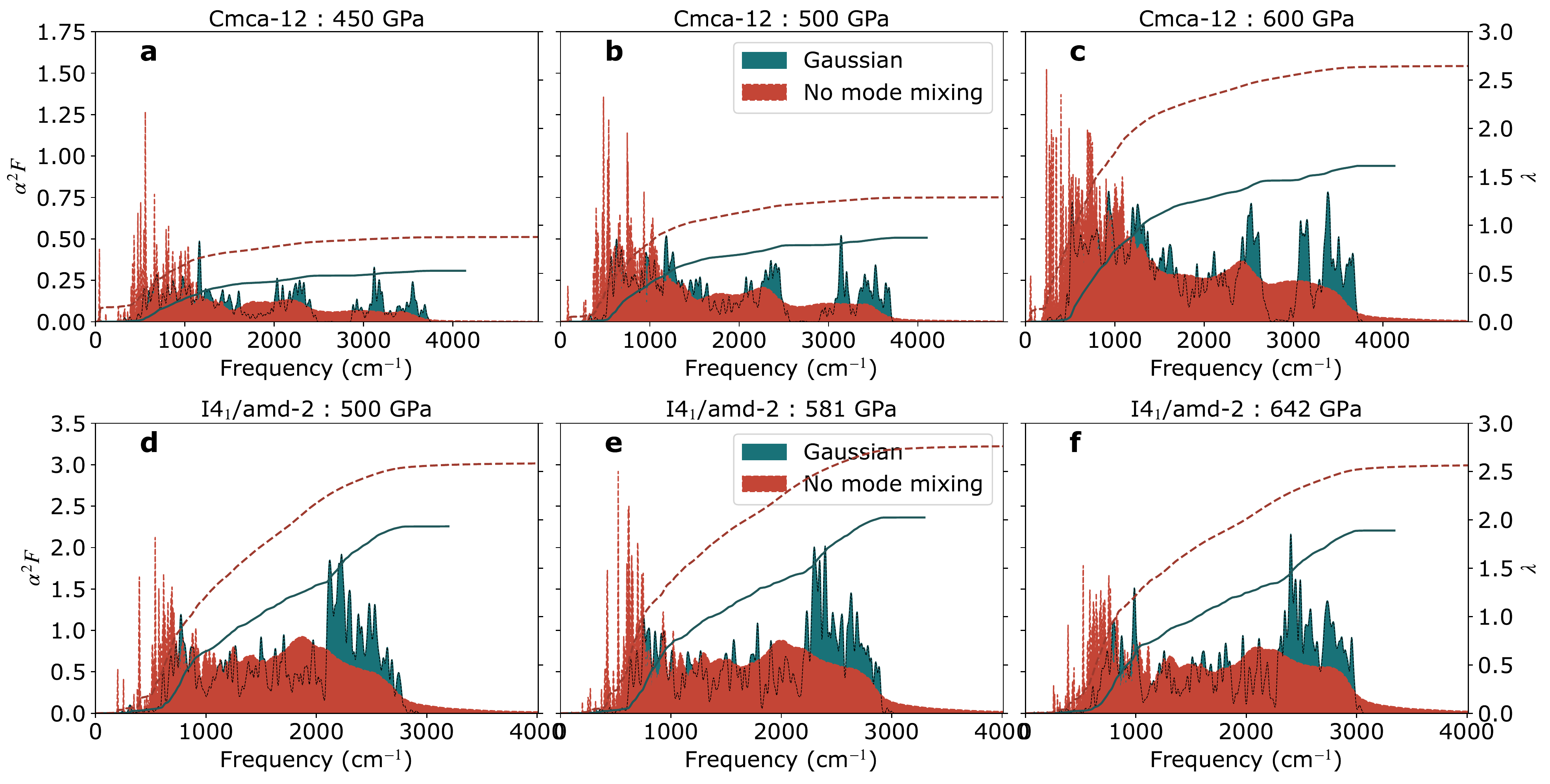}
			\caption{Eliashberg spectral function and electron-phonon coupling constant of solid hydrogen in molecular Cmca-12 phase VI at (a) 450 GPa, (b) 500 GPa, (c) 600 GPa and atomic tetragonal I4$_1$/amd-2 phase at (d) 500 GPa, (e) 581 GPa and (f) 642 GPa. Gaussian represents results where we used SSCHA structures and second order force constants (auxiliary force constants), but calculated $\alpha ^2F$ by using Eq.~\ref{eq_dirac} for the definition of spectral function (smearing in Gaussian function of 10 cm$^{-1}$). No mode mixing represents results where we used SSCHA structures and second-order force constants (auxiliary force constants), but calculated $\alpha ^2F$ by using Eq.~\ref{eq_full} for the definition of the spectral function.}
			\label{supfig3}
		\end{center}
	\end{figure*}
	
	In Supp. Figure~\ref{supfig3} we compare results for the Eliashberg spectral function $\alpha ^2F (\omega)$ calculated using Eq.~\ref{eq_dirac} but with phonon frequencies and polarization vectors coming from the SSCHA auxiliary force constants, not the harmonic phonons, (labeled Gaussian) and Eq.~\ref{eq_full} with the phonon self-energy from Eq.~\ref{eq_self_energy} (labeled Anharmonic). This Gaussian approach is the one that has been used so far in the literature to estimate the anharmonic renormalization of the Eliashberg function. Thus, in both cases, we used properties obtained with SSCHA (structure and interatomic force constants). Treating phonon spectral function in the dynamical bubble approximation further softens phonon modes. In molecular phase VI, this leads to the complete closing of the vibron-phonon gap. This softening mostly increases the final electron-phonon coupling strength. Additionally, in the ``no mode mixing'' calculation we can see a longer tail at higher frequencies which is a consequence of the phonon-phonon interaction that includes the annihilation of two phonon modes. 
	
\begin{figure*}[!t]
		\begin{center}
			\includegraphics[width=0.9\textwidth]{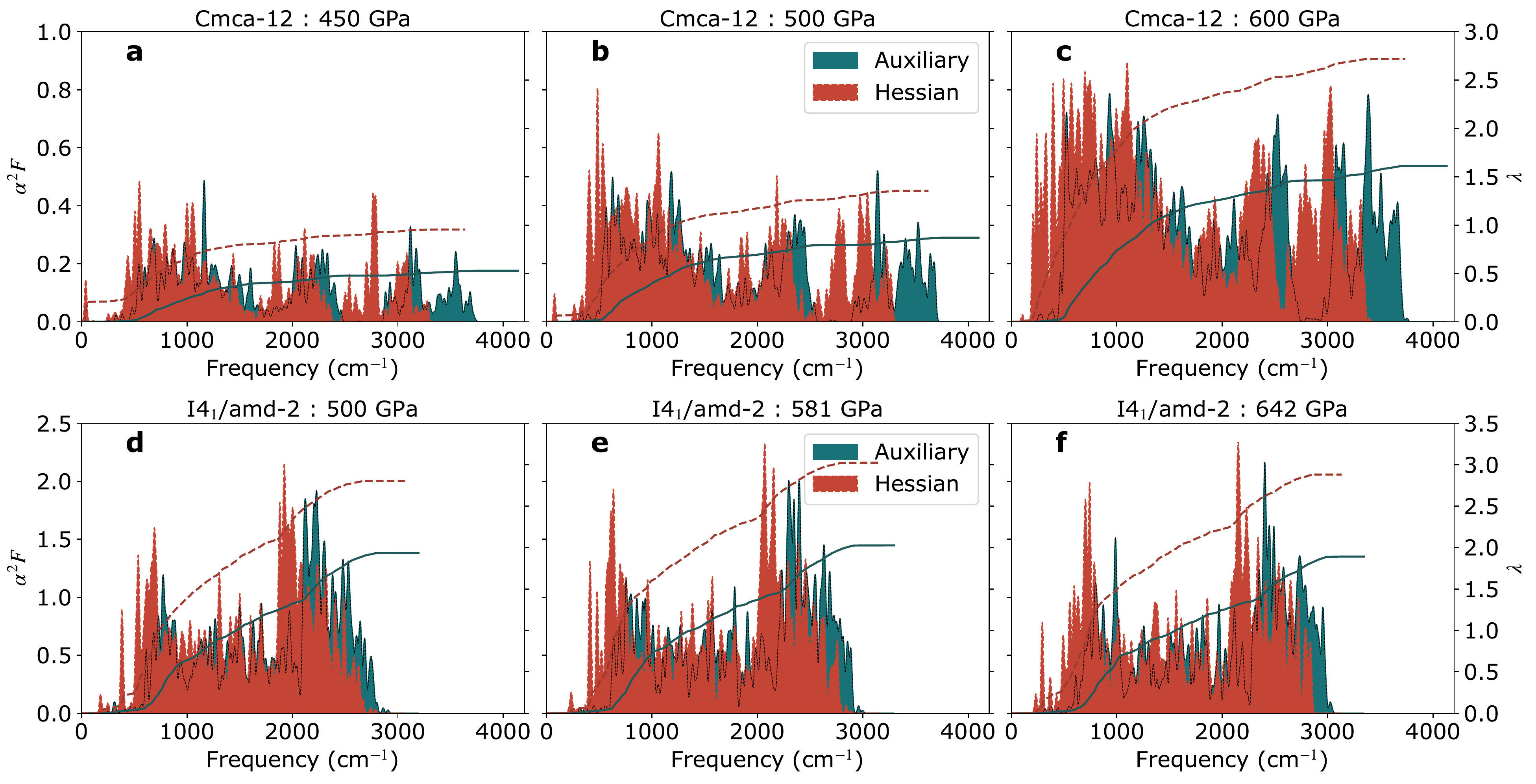}
			\caption{Eliashberg spectral function and electron-phonon coupling constant of solid hydrogen in molecular Cmca-12 phase VI at (a) 450 GPa, (b) 500 GPa, (c) 600 GPa and atomic tetragonal I4$_1$/amd-2 phase at (d) 500 GPa, (e) 581 GPa and (f) 642 GPa. Auxiliary represents results where we used SSCHA structures and second order force constants (auxiliary force constants) but calculated $\alpha ^2F$ by using Eq.~\ref{eq_dirac} for the definition of spectral function (smearing in Gaussian function of 10 cm$^{-1}$). Hessian represents results where we used SSCHA structures and eigenvalues from free energy Hessian and calculated $\alpha ^2F$ by using Eq.~\ref{eq_dirac} for the definition of the spectral function.}
			\label{supfig5}
		\end{center}
	\end{figure*}

	To gauge the influence of this change of the Eliashberg spectral function on the superconductivity we calculated critical temperature using the isotropic Migdal-Eliashberg equation. Results for different calculations are shown in Supp. Table~\ref{tb1}. The ``Anharmonic'' calculations consistently show higher critical temperatures in both phases. This is mainly due to the softening of the phonon modes due to the anharmonicity. Additionally, we calculate the Eliashberg spectral function in Gaussian approximation using free energy Hessian phonons, see Supp. Fig.~\ref{supfig5}. These calculations consistently give higher critical temperature which is again a consequence of the softening of the phonon modes in this approximation. Calculations with Hessian phonons give larger T$_{\mathrm{C}}$s compared to calculations done with realistic broadening in no mode mixing approximation. This points to the conclusion that the softening of the phonon modes is beneficial for the superconductivity, while the broadening of the spectral lines is disruptive. 
	
	Further, we can discuss the hierarchy of influences on the estimation of the critical temperature. In the molecular phase, we find that the SSCHA renormalization of structure and phonons has a larger effect on the critical temperature than the inclusion of the realistic phonon broadening. On the other hand, in the atomic phase of hydrogen, we find the opposite effect. In the molecular phase, the enhancement of critical temperature estimate comes mainly from the increase of the electronic density of states at the Fermi level due to SSCHA structural renormalization. In the atomic phase, the change in the density of states does not have a large impact on critical temperature and because of this dynamical renormalization of phonons has a larger effect. 
	
	From the mode projected spectral functions $\sigma_{\mu}(\mathbf{q}, \omega)$ one can get the Cartesian based spectral function with $\sigma_{ab}(\mathbf{q},\omega)=\sum_{\mu}e^a_{\mu}(\mathbf{q})e^{b*}_{\mu}(\mathbf{q}) \sigma_{\mu}(\mathbf{q},\omega)$, where $e_{\mu}(\mathbf{q})$ is the eigenvector of the phonon with branch $\mu$ and wave vector $\mathbf{q}$. However, one can avoid making the no-mode mixing approximation, which is made in Eq.~\ref{eq_self_energy}, by calculating the full matrix of the phonon self-energy: 
	\begin{align*}
		\Pi_{\mu\mu '}(\mathbf{q}, \omega) &= \frac{1}{N_\mathbf{k}}\sum _{\mathbf{k}_{1},\nu}\sum _{\mathbf{k}_{2},\rho}\sum _{\mathbf{G}}\delta _{\mathbf{G}, \mathbf{q} + \mathbf{k}_{1} + \mathbf{k}_{2}}D_{\mathbf{q}\mathbf{k}_{1}\mathbf{k}_{2}}^{\mu\nu\rho}D_{-\mathbf{k}_{1}-\mathbf{k}_{2}-\mathbf{q}}^{\nu\rho\mu '}\times \\
		&\times\frac{\hbar}{4\omega _{\nu}(\mathbf{k}_{1})\omega _{\rho}(\mathbf{k}_{2})}\left(\frac{(\omega _{\nu}(\mathbf{k}_{1})-\omega _{\rho}(\mathbf{k}_{2}))(n _{\nu}(\mathbf{k}_{1}) - n _{\rho}(\mathbf{k}_{2}))}{(\omega _{\nu}(\mathbf{k}_{1})-\omega _{\rho}(\mathbf{k}_{2}))^2 - \omega ^2} - \frac{(\omega _{\nu}(\mathbf{k}_{1})+\omega _{\rho}(\mathbf{k}_{2}))(n _{\nu}(\mathbf{k}_{1}) + n _{\rho}(\mathbf{k}_{2}) + 1)}{(\omega _{\nu}(\mathbf{k}_{1})+\omega _{\rho}(\mathbf{k}_{2}))^2 - \omega ^2}\right). \numberthis
		\label{eq_self_energy1}
	\end{align*}
	Plugging this self-energy in the Dyson equation one can get a complete phonon Green's function:
	\begin{align*}
		G _{\mu\mu'}(\mathbf{q}, \omega) = \left[\omega ^2\delta_{\mu\mu'} - D _{\mu\mu'}(\mathbf{q}) - \Pi _{\mu\mu'}(\mathbf{q} ,\omega)\right]^{-1}.
	\end{align*}
	Here $D _{\mu\mu'}(\mathbf{q})$ is the auxiliary SSCHA dynamical matrix in the mode basis (it is diagonal in this basis). Since we are taking the inverse in the equation above, we in essence are mixing self-energies of different phonon branches. The full phonon spectral function is then straightforwardly calculated as $\sigma_{\mu\mu'}(\mathbf{q}, \omega) = -\frac{\omega}{\pi}\mathrm{Im}G _{\mu\mu'}(\mathbf{q}, \omega)$, which can later be projected onto the Cartesian basis:
	\begin{align}
		\sigma_{ab}(\mathbf{q},\omega)=\sum_{\mu,\mu'}e^a_{\mu}(\mathbf{q})e^{b*}_{\mu'}(\mathbf{q}) \sigma_{\mu\mu'}(\mathbf{q},\omega).
		\label{eq_mode_mixing}
	\end{align} 
	
	Finally, we calculate the critical temperature in the atomic phase using the full spectral function, see Supp. Fig.~\ref{supfig8}. There are only small differences between these results and the ones in no mode mixing approximation. Comparing the phonon density of states in these two approaches (see Supp. Fig.~\ref{supfig6}) we see that these are almost identical explaining the similarity between no mode mixing and full spectral function approaches. 
	
	\begin{figure}
		\centering
		\includegraphics[width=0.9\textwidth]{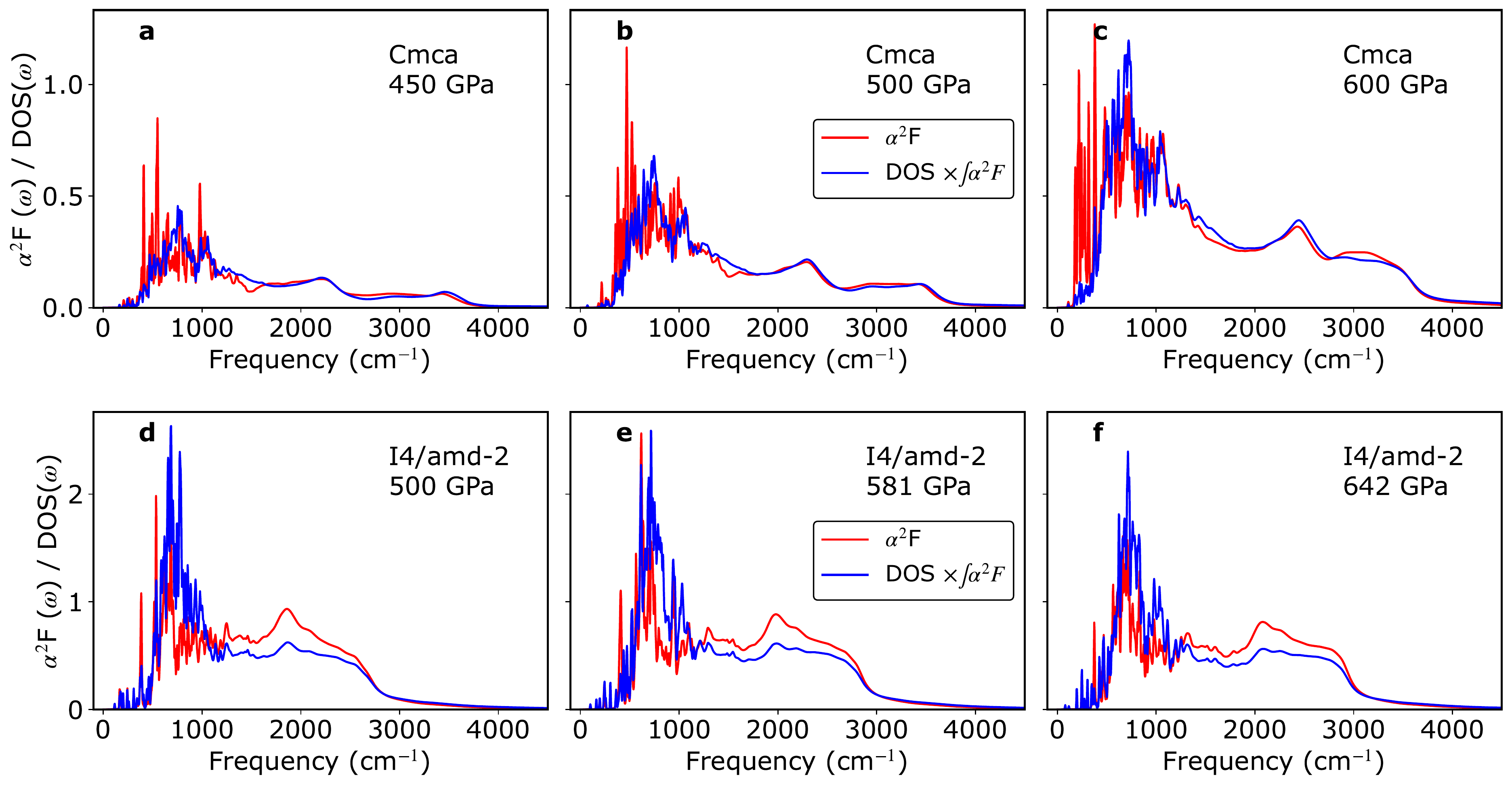}
		\caption{Comparison between Eliashberg spectral function $\alpha ^2F$ and phonon density of states both calculated with the full phonon spectral functions for the molecular Cmca phase of solid hydrogen at a) 450 GPa, b) 500 GPa, c) 600 GPa, and the atomic phase of solid hydrogen at d) 500 GPa, e) 581 GPa and f) 642 GPa. Phonon densities of states were scaled by the integral of the Eliashberg spectral function for better comparison.}
		\label{fig:a2f_dos}
	\end{figure}
	
	\begin{figure*}[!t]
		\begin{center}
			\includegraphics[width=0.95\textwidth]{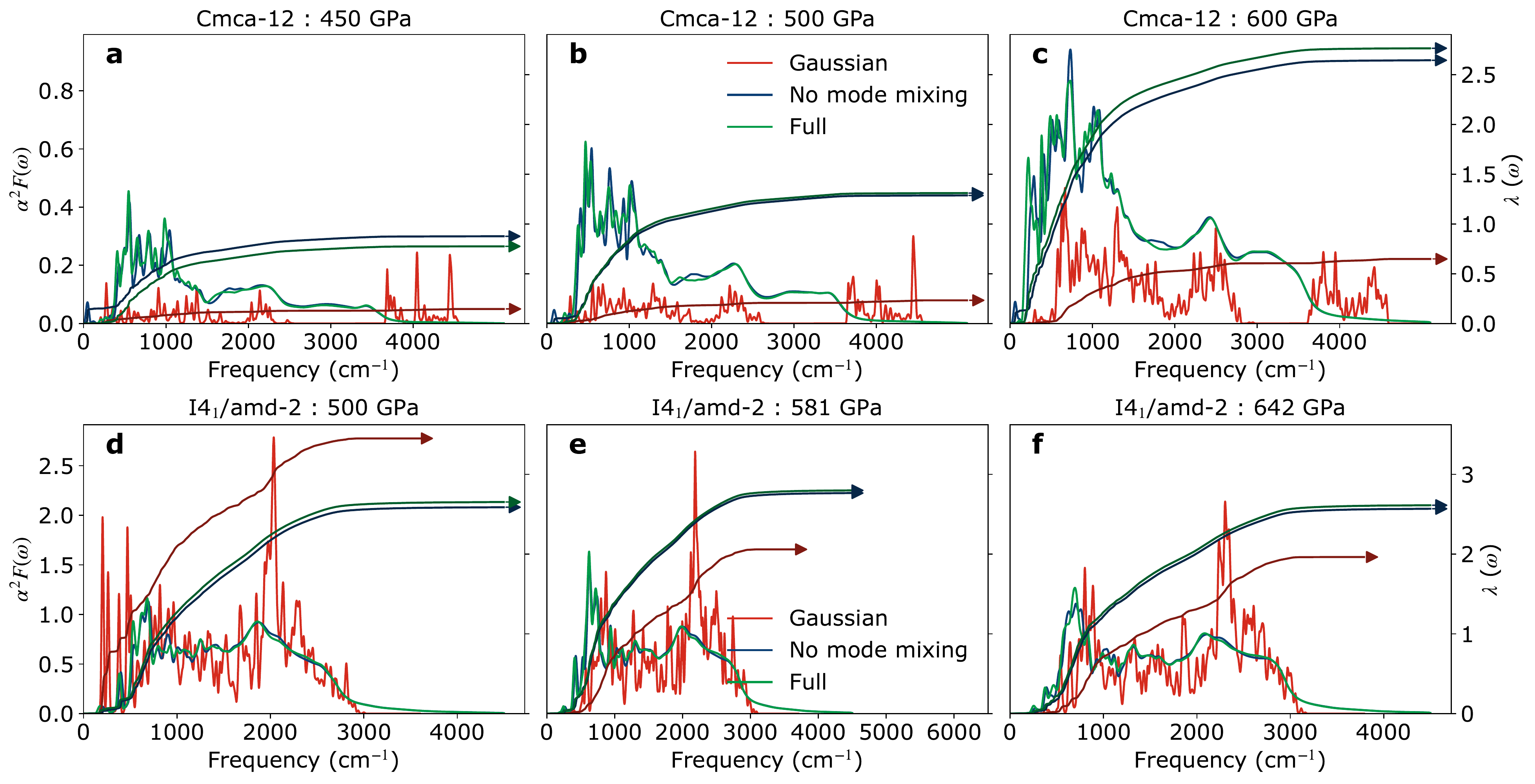}
			\caption{Eliashberg spectral function $\alpha^2F(\omega)$ and integrated electron-phonon coupling constant $\lambda(\omega)$ of solid hydrogen in molecular Cmca-12 phase VI at (a) 450 GPa, (b) 500 GPa, (c) 600 GPa, and atomic tetragonal I4$_1$/amd-2 phase at (d) 500 GPa, (e) 581 GPa, and (f) 642 GPa calculated in the SSCHA using the spectral function calculated fully and in the no mode mixing approximation, and in the harmonic case using Gaussian method.}
			\label{supfig8}
		\end{center}
	\end{figure*}
	
	\begin{figure*}[!t]
		\begin{center}
			\includegraphics[width=0.95\textwidth]{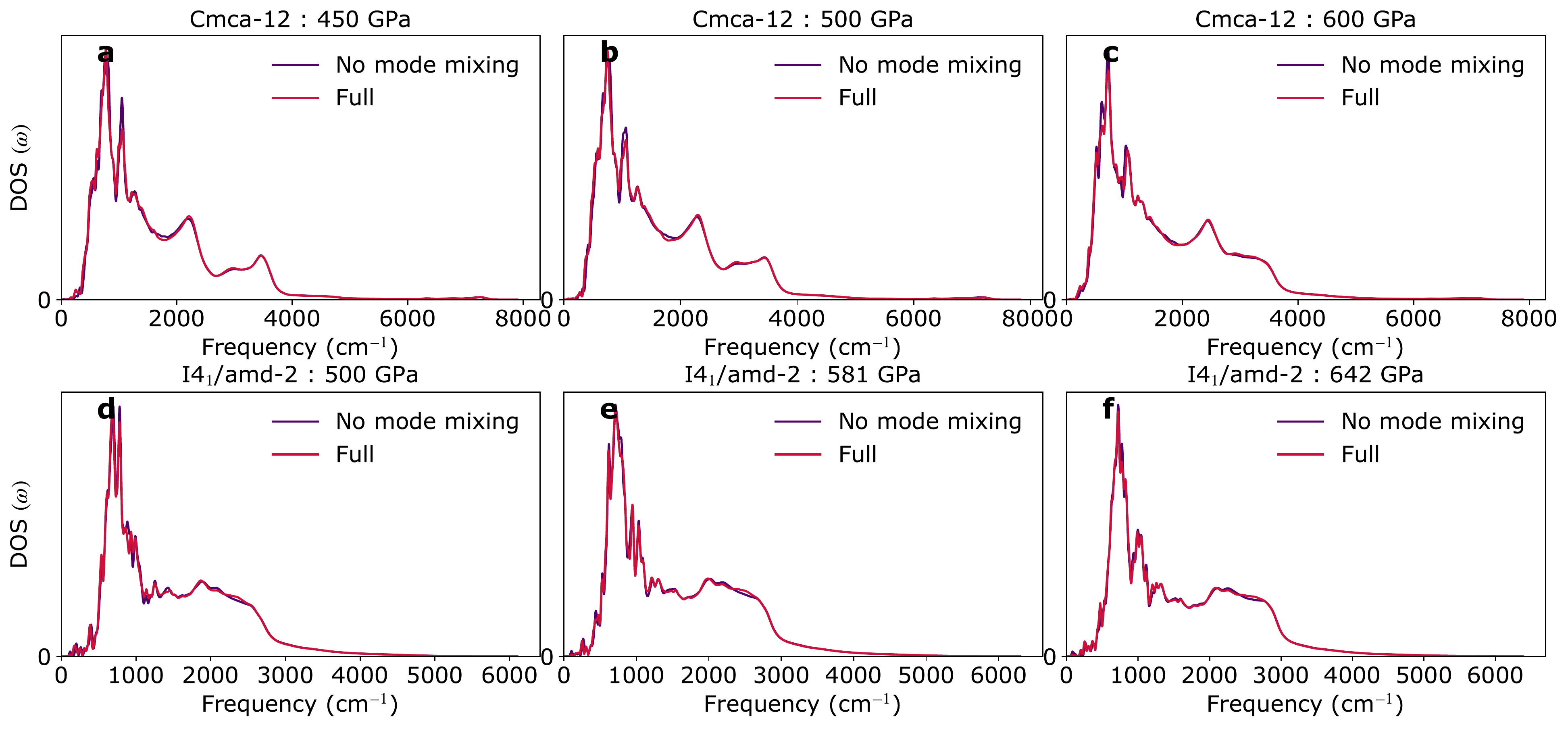}
			\caption{Phonon density of states calculated with the full phonon spectral functions and with the ones obtained in the no mode mixing approximation for the molecular Cmca phase of solid hydrogen at a) 450 GPa, b) 500 GPa, c) 600 GPa, and the atomic phase of solid hydrogen at d) 500 GPa, e) 581 GPa and f) 642 GPa.}
			\label{supfig6}
		\end{center}
	\end{figure*}
	
	\begin{table}[h]
		\begin{tabular}{|c|c|c|c||c|c|c|}
			\hline
			T$_\mathrm{C}$	 & \multicolumn{3}{c||}{molecular Cmca-12} & \multicolumn{3}{c|}{atomic I4$_1$/amd-2} \\
			\hline \hline
			& 450 GPa & 500 GPa & 600 GPa & 500 GPa & 581 GPa & 642 GPa      \\
			\hline
			Gaussian - Harmonic                 &  0 K  & 0 K     & 35 K    & 349 K   & 330 K   & 321 K \\
			\hline
			Gaussian - Auxiliary                &  19 K & 83 K    & 219 K   & 313 K   & 334 K   & 322 K \\
			\hline
			Gaussian - Hessian                  &  52 K & 129 K  & 268 K   & 356 K   & 376 K   & 370 K \\
			\hline
			No mode mixing    &  45 K & 119 K  & 252 K   & 336 K   & 356 K   & 347 K \\
			\hline
			Full &   50 K & 123 K & 254 K   & 337 K   & 356 K   & 347 K \\
			\hline
		\end{tabular}
		\caption{Superconducting critical temperature in solid hydrogen molecular VI Cmca-12 and atomic I4$_1$/amd-2 phases. Harmonic results are obtained using DFT structures, harmonic phonons with Eliashberg spectral function calculated using Eq.~\ref{eq_dirac}. Auxiliary/Hessian results are obtained using SSCHA structures, and SSCHA auxiliary/hessian phonons with Eliashberg spectral function calculated using Eq.~\ref{eq_dirac}. No mode mixing and full results are obtained using SSCHA structures, SSCHA auxiliary phonons with Eliashberg spectral function calculated using Eq.~\ref{eq_full} and Eq.~\ref{eq_self_energy1}. All of these results are obtained using the same exchange-correlation functional.}
		\label{tb1}
	\end{table}
	
	To check which phonon modes contribute significantly to electron-phonon coupling, we compared the Eliashberg spectral function and phonon density of states calculated with full phonon spectral functions in Supp. Fig.~\ref{fig:a2f_dos}. In the molecular $Cmca$ phase coupling mostly comes from low-frequency phonons, while in the atomic phase, the higher-frequency phonon contributes more significantly. However, in both cases, there is no dominant frequency range that contributes the most to the total electron-phonon coupling constant.
	
\begin{figure*}[h]
		\begin{center}
			\includegraphics[width=0.45\textwidth]{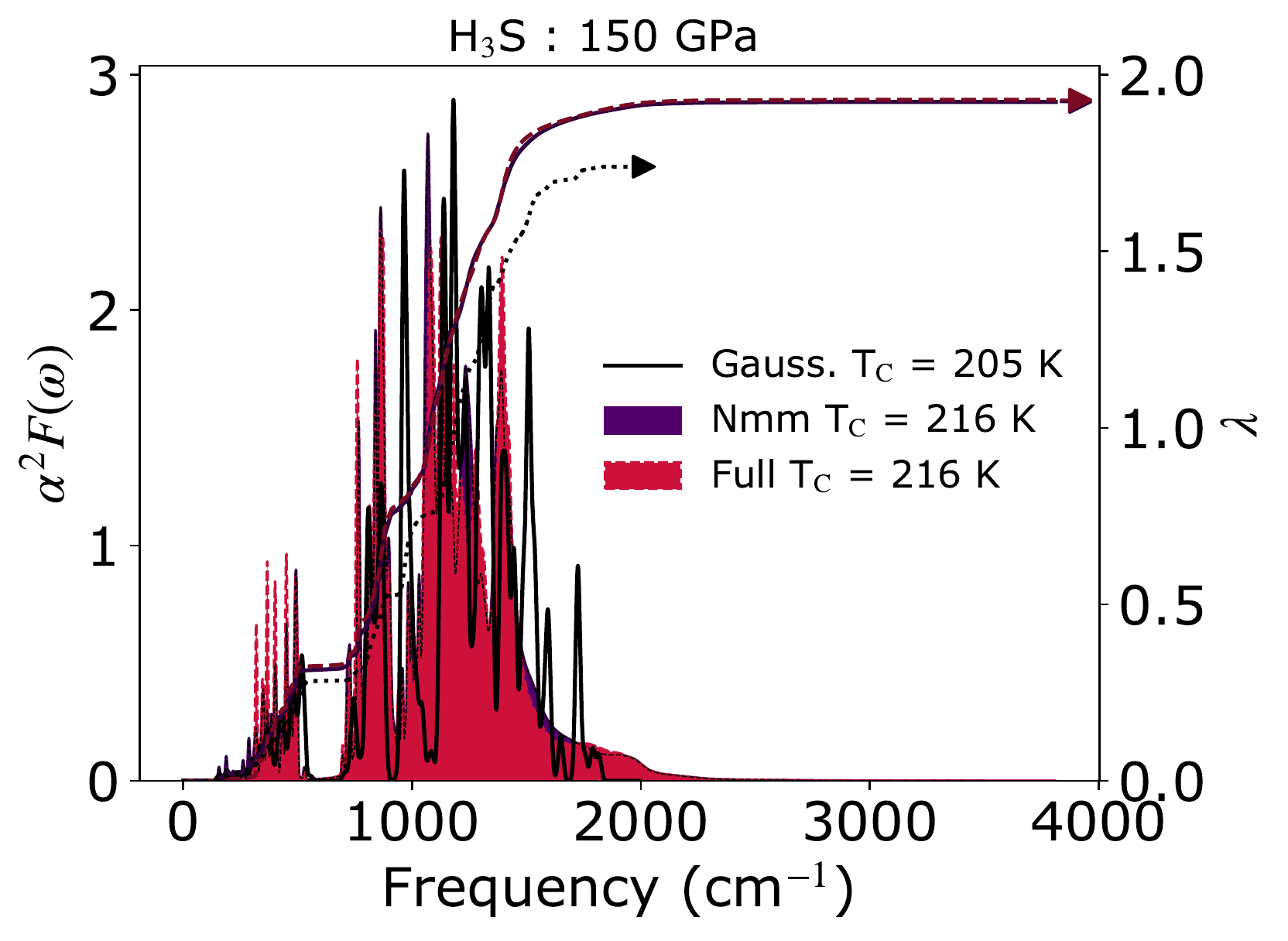}
			\caption{Eliashberg spectral function of H$_3$S at 150 GPa calculated with three different methods. "Gauss." stands for Gaussian approximation (Eq.~\ref{eq_dirac}), "Nmm" is the no mode mixing approximation (Eq.~\ref{eq_full}), and "full" is the calculation with full phonon spectral function (Eq.~\ref{eq_mode_mixing}).}
			\label{supfig7}
		\end{center}
	\end{figure*}

	To make sure that these results are reliable, we also calculated the critical temperature of H$_3$S at 150 GPa using this method. The simple Gaussian approximation already gives a very good estimation of the critical temperature in this system and the new approach including phonon spectral functions should not change it \cite{IonHS2}. This is what we see in our calculations, where anharmonic phonon spectral functions have a limited effect on critical temperature, see Supp. Fig.~\ref{supfig7}.
	
	\subsection{Convergence studies}
	
\begin{figure}
		\centering
		\includegraphics[width=0.9\textwidth]{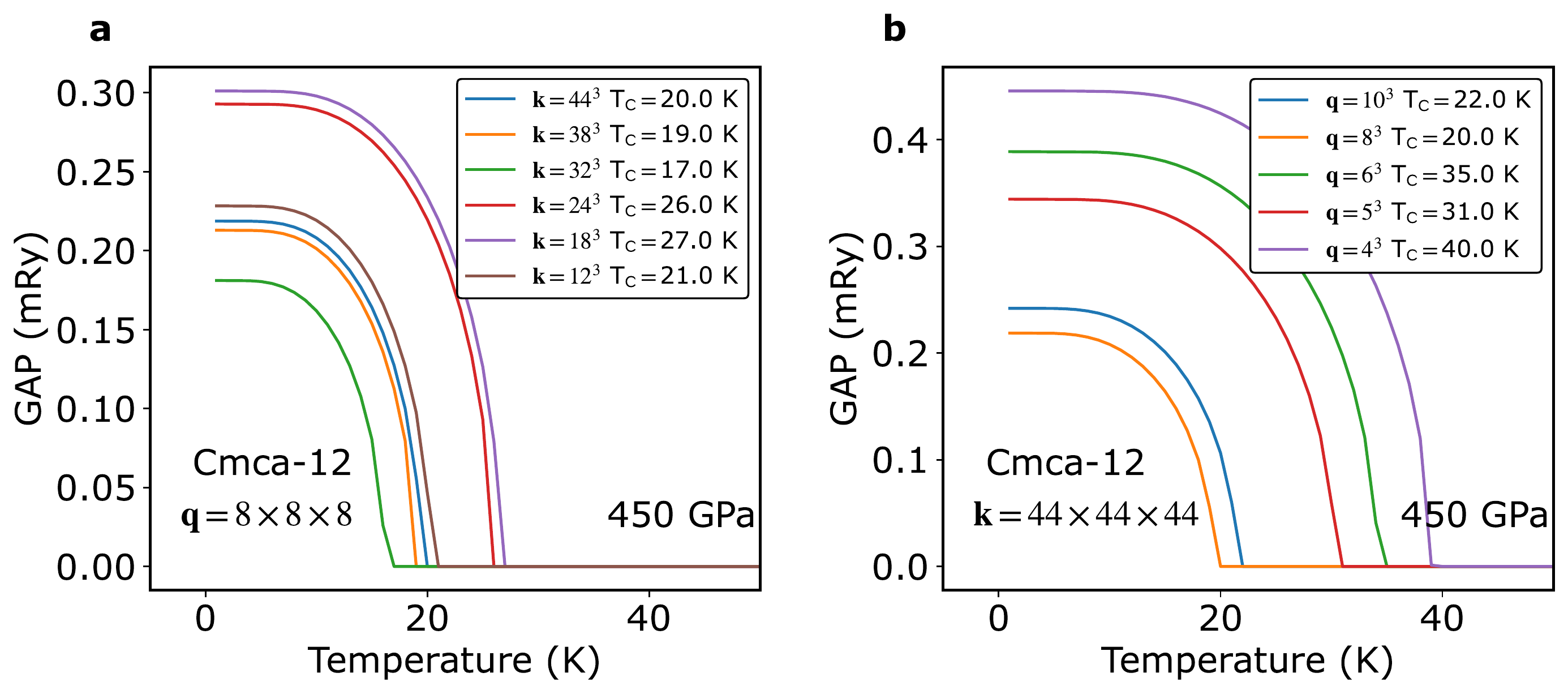}
		\caption{Convergence study of the critical temperature in molecular VI phase of hydrogen at 450 GPa with respect to \textbf{a} \textbf{k} point grid and \textbf{b} \textbf{q} point grid. Convergence study of the critical temperature in the atomic phase of hydrogen at 642 GPa with respect to \textbf{c} \textbf{k} point grid and \textbf{d} \textbf{q} point grid.}
		\label{convergence}
	\end{figure}

	We have performed convergence studies of critical temperature with respect to $k$-point and $q$-point grids, as shown in Supp. Fig.~\ref{convergence}.
	
\begin{figure}
		\centering
		\includegraphics[width=0.9\textwidth]{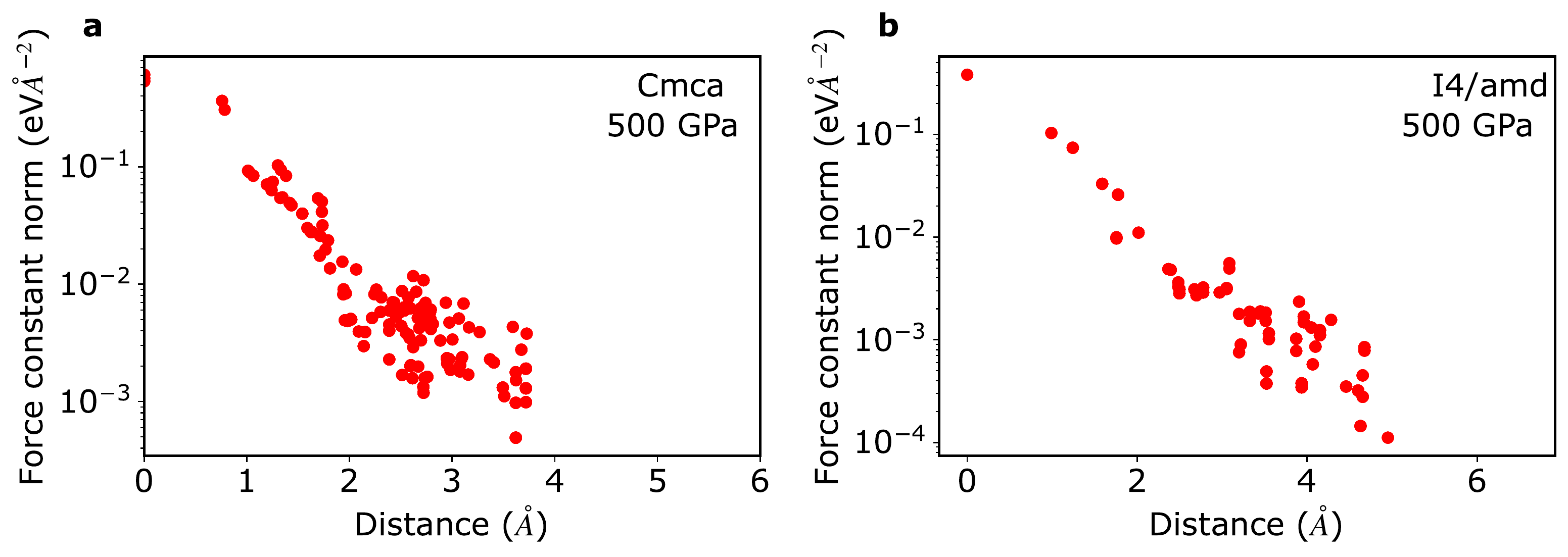}
		\caption{Decay of the second order force constant with atom-atom distance for \textbf{a} molecular phase and \textbf{b} atomic phase of solid hydrogen at 500 GPa.}
		\label{decay}
	\end{figure}

	The size of the system precludes us from checking the convergence of results with the size of the SSCHA supercell. However, in Supp. Fig.~\ref{decay} we are showing the decay of the second and third-order force constants to justify the use of the interpolation method in obtaining vibrational properties of solid hydrogen.
	
\end{document}